\documentclass[12pt,a4paper]{article}
\usepackage{jheppub}
\usepackage[T1]{fontenc}
\usepackage{slashed}
\usepackage{caption}
\usepackage{subcaption} 
\usepackage{float}

\usepackage{empheq}%

\usepackage{comment}

\usepackage{xcolor}
\usepackage{color,soul}

\usepackage{bbold}

\usepackage{graphicx}

\usepackage{feynmp}

\def\Beq{\begin{equation}}
\def\Eeq{\end{equation}}

\def\Bal{\begin{aligned}}
\def\Eal{\end{aligned}}

\def\[{\left[}
\def\]{\right]}
\def\({\left(}
\def\){\right)}

\newcommand{\be}{\beta}

\def \be {\begin{equation}}
\def \ee {\end{equation}}
\def \bea {\begin{eqnarray}}
\def \eea {\end{eqnarray}}
\def \beal#1 {\begin{align}#1\end{align}}

\everymath{\displaystyle}

\preprint{}

\title{Thermal Bekenstein-Hawking entropy from the worldsheet }
 \author[]{Indranil
  Halder\note{ihalder@g.harvard.edu} and}  \author[]{Daniel L. Jafferis\note{jafferis@g.harvard.edu}}

\affiliation{Jefferson Physical Laboratory, Harvard University, Cambridge, MA 02138, USA}

\abstract{
We define and compute the leading sphere diagram contribution to the entropy of the BTZ black hole supported by Kalb-Ramond flux in bosonic string theory. In a winding condensate description, integrating exactly over the constant mode for the radial direction of AdS$_3$ reduces the problem to one of the correlation functions of winding operators in the free theory. The volume of the residual PSL(2,$\mathbb{C}$) gauge group of the sphere is canceled by the action of conformal transformations on the winding interaction insertions. We formulate a precise version of the replica trick in terms of (infinitesimally) non-integer winding condensates to produce the entropy of the BTZ black hole. The resulting entropy can be calculated from the one-point function of a non-local operator on the worldsheet. 
}

\begin{document}

\maketitle

\section{Introduction}

Entropies in gravitational theories have a deep significance as their leading order value in weakly gravitating system is encoded in classical geometry via the Bekenstein-Hawking \cite{PhysRevD.7.2333, Hawking:1975vcx} and Hubeny-Rangamani-Ryu-Takayanagi \cite{Ryu:2006bv, Hubeny:2007xt} area formulas. These can be derived in the long-distance effective theory, explaining their universality, using thermodynamic relations such as the Euclidean free energy \cite{PhysRevLett.28.1082, PhysRevD.15.2752}\footnote{Alternatively, the Bekenstein-Hawking entropy can be calculated using the method of orbifolds \cite{Dabholkar:1994ai, Dabholkar:2001if}.   Further gravitational loop corrections were found Sen \cite{Sen:2008vm}.} and the replica trick \cite{Lewkowycz:2013nqa} \footnote{In this paper we need only the leading order in small $g_s$ version of Lewkowycz-Maldacena \cite{Lewkowycz:2013nqa}. The first quantum correction was found by  Faulkner-Lewkowycz-Maldacena \cite{Faulkner:2013ana}. For further sub-leading order corrections and current status see the work of Engelhardt-Wall \cite{Engelhardt:2014gca}.}calculations. It is therefore of great interest to find the string theory generalization of the area-entropy relation. This is the question we will address for the BTZ black hole - what is the $\alpha'$ exact analog of the area operator on the worldsheet?\footnote{More precisely we are looking at string compactifications of the type BTZ$\times$M, where M is an compact internal manifold. To have a tractable description of the worldsheet in the NSR formalism we will consider the bosonic string analog of the setup in \cite{Maldacena:1998bw} with pure NS-NS flux. }

There are two conceptual hurdles to overcome in attempting to perform the Gibbons-Hawking-York Euclidean black hole calculation of the entropy in string theory. First, our interest is in the leading term in the (logarithm of the) partition function, of order $g_s^{-2}$, which comes from the genus zero diagram with no insertions. This corresponds to the on-shell action of the target space theory. This sphere diagram is notoriously difficult to define, due to the unfixed PSL(2,$\mathbb{C}$) gauge group\footnote{In a beautiful recent work \cite{Eberhardt:2023lwd}, the zero temperature partition function in AdS$_3$ is analyzed. In their approach, a new parameter $\mu$ is introduced in the worldsheet CFT. The triple derivative of the partition function with respect to $\mu$ is studied, thereby avoiding the issue of unfixed gauge modes. } (more so because there is no natural regularization that assigns a finite volume to  PSL(2,$\mathbb{C}$)\footnote{The partition function on the disc is also plagued by similar issues of unfixed gauge modes. In this case, Polchinski showed how to get a sensible finite answer for the volume of the residual unfixed gauge group using a suitable regularization  \cite{Liu:1987nz}. For a re-derivation of D-brane tension \cite{Polchinski:1995mt} using this result see \cite{Eberhardt:2021ynh}.}) and the infinite volume of the zero mode integration over the target space.\footnote{The problem of unfixed gauge degrees of freedom also appears in on-shell two or lower-point amplitudes of string theory. In the context of the two-point amplitude in flat-space string theory, the authors of \cite{Erbin:2019uiz} showed that the divergence from the unfixed gauge modes is canceled by the IR divergence of zero modes in target space to produce a sensible result (a similar conceptual point was speculated in \cite{Troost} for the one-point function of the central charge operator in AdS$_3$. A non-trivial $l_{AdS}/l_s=\sqrt{k-2}$ dependent map between the IR cut-off in the target space and the regulator used to define the volume of PSL(2,$\mathbb{C}$) is proposed).} Furthermore, string theory, even in its formulation as string field theory, is defined with respect to a background, so the only well-posed question appears to be the difference in the on-shell action of one background with respect to another.

We resolve these issues by using an exact dual description of the BTZ worldsheet theory in terms of an Euclidean time winding condensate on a free field background \cite{Jafferis:2021ywg, Halder:2022ykw,Halder:2023nlp}. This description is strongly coupled in the large $k$ limit, however, we can perform the calculations exactly. Note that these backgrounds have the same asymptotics, so they differ by a finite normalizable deformation (assuming the length of AdS in the strings unit is large\footnote{More precisely we are considering $k>3$.}).
The free linear dilaton background provides the reference point for the genus zero calculation, which can be performed using the standard technique of analytic continuation away values with integer numbers of insertions of the winding interaction vertex operators. The residual conformal transformations can now be fixed by localizing three of those insertions. 

The second conceptual point is that the result of the above computation in fact vanishes. This corresponds to the fact that the bulk on-shell action of string field theory is zero \cite{Erler:2022agw} by the dilaton theorem  \cite{Bergman:1994qq} (considerations of a  boundary in spacetime is much more subtle \cite{Adams:2001sv, Dabholkar:2001if, Okawa:2004rh}, see \cite{Headrick:2004hz} for a review), and the our worldsheet calculation does not add any boundary terms (like the Gibbons-Hawking-York term, these boundary terms are responsible for the nonzero result in spacetime gravity, the Einstein-Hilbert bulk term can be written (up to total derivatives) in terms of the $\beta$ functions of the worldsheet sigma model which vanishes on-shell to all orders in $\alpha' $ \cite{Tseytlin:1988tv, Tseytlin:1988rr, Osborn:1989bu, Osborn:1991gm}). 

Therefore, we turn to the Lewkowycz-Maldacena method of computing the entropy from analytic extrapolation of the $Z_n$ quotient of the $n^{\textrm{th}}$ Renyi geometries, which only requires knowledge of the bulk action. In semi-classical gravity, the result is that the entropy can be extracted from the change in the bulk action upon the addition of an infinitesimal conical defect in the interior  \cite{Carlip:1993sa, Banados:1993qp, Susskind:1994sm,  Fursaev:2006ih, Headrick:2010zt} (see also \cite{Brustein:2021cza}), keeping fixed the asymptotic geometry. The string theory analog is an analytic continuation of the worldsheet theories corresponding to a $Z_n$ orbifold in the interior and the same Euclidean black hole asymptotics. 

This is somewhat related to the proposal of Susskind-Uglum \cite{Susskind:1994sm}, utilizing the off-shell worldsheet formalism of  \cite{Tseytlin:1988tv, Tseytlin:1988rr}. Here one is instructed to take a derivative of the off-shell worldsheet non-linear sigma model partition function with respect to the renormalization group UV cut-off to produce the string theoretic partition function (for a recent discussion of it see \cite{Ahmadain:2022tew, Ahmadain:2022eso}). The method relies on the details of the perturbative renormalization group flow and the simplest way of performing the path integral produces the bulk term in the target space, order by order in $\alpha'$ expansion. In this paper, we will produce exact in $\alpha'$ results without dealing with renormalization group flow.

The conical deformation we will use corresponds, in the winding condensate description, to a background obtained by adding a non-integer winding operator to the free theory. Such operators are not mutually local with the properly quantized Matsubara frequency vertex operators, but they are local amongst themselves, so the sphere diagram remains well-defined. Moreover, this conical defect-inducing operator is weight $(1,1)$ in the free worldsheet theory.

One might worry that such a $(1,1)$ deformation would correspond to the inclusion of a physical source for the curvature of the conical defect, which would result in a cancelation of its contribution to the action (at the linear order relevant for the entropy). However, the non-integer winding deformation that we use precisely does not include the reflected wave component in the radial direction. Including the reflected wave would result in a deformation that was not normalizable at the AdS boundary, and not including it corresponds to taking a singular conical bulk geometry without a compensating brane, as desired. 

The worldsheet sphere amplitude associated with this deformed background is non-vanishing, and we compute it with the same free field 
methodology. From this, we find the $g_s^{-2}$ leading order contribution to the BTZ entropy in string theory, exact in $\alpha'$, which in $\alpha'\to 0$ limit reproduces the expected result. The derivative of the winding operator with respect to the non-integer replica number gives a non-local worldsheet operator that is the stringy generalization of the area.

Beyond the leading order in string coupling the thermal entropy gets a contribution both from the one-point function of the stringy area operator mentioned above and the entanglement entropy of the graviton and the other bulk fields outside the horizon. A systematic discussion of the finite value of entanglement entropy of the bulk fields at one loop including all stringy corrections can be obtained by the `extrapolated' (order of the orbifold is fractional) Dabholkar-Witten orbifold method in \cite{Witten:2018xfj, Dabholkar:2022mxo} (see also \cite{He:2014gva}).

The paper is organized as follows. In section \ref{2} we review the sigma model and the winding condensate description of the worldsheet CFT. Then in section \ref{3} we calculate the sphere path integral for BTZ carefully separating the divergent and finite parts. In section \ref{4} we formulate the replica trick on the worldsheet. We conclude by discussing several open questions in section \ref{5}.

\section{Review of the duality}\label{2}

In this paper, we will consider bosonic strings on AdS$_3\times$M, M is a compact manifold, supported purely by Kalb-Ramond flux (the left and right central charges of the holographic dual CFT$_2$ are taken equal to each other). We will focus our discussion on the worldsheet CFT for the AdS$_3$ part. In subsection 2.1 we review the standard description of the worldsheet CFT in terms of WZW model. In subsection 2.2 we review the winding condensate description developed in \cite{Jafferis:2021ywg, Halder:2022ykw}. Throughout this section, we will follow the notation and the conventions used in \cite{Halder:2022ykw}.

\subsection{Sigma model description}\label{2.1}

The usual sigma model description of the worldsheet CFT in Poincare patch\footnote{The metric in our convention is given by
$ds^2=l_{AdS}^2(d\hat{\phi}^2+e^{2\hat{\phi}}d\hat{\Gamma}d\hat{\bar{\Gamma}})$.
} takes the following form
\begin{equation}\label{sigma_model}
	\begin{aligned}
		& S_{\text{sigma}}=\frac{k}{2 \pi}\int 2 d^2 \sigma \(\partial \hat{\phi} \bar{\partial} \hat{\phi} +\bar{\partial} \hat{\Gamma} \partial \bar{\hat{\Gamma}}e^{2\hat{\phi}} \)\\
	\end{aligned}
\end{equation}
The Kalb-Ramond flux $k$ determines the radius of AdS in string units
\begin{equation}
    l_{AdS}=\sqrt{k-2}l_s
\end{equation}
For the rest of the paper, unless otherwise noted, we choose to work with the units set by $l_s=1$. The sigma model description in (\ref{sigma_model}) is exact at leading order in the large $k$ limit. It is well known that for finite values of $k$, the (Euclidean target) worldsheet CFT is given by the SL(2,$\mathbb{C}$)$_k$/SU(2) WZW model. 

The most convenient way to represent the simplest set of vertex operators in the large $k$ limit is the following:\footnote {There is a sign different between the convention here and in \cite{Ribault:2005wp}: $\Phi_{-1-j}(\text{here})=-\Phi_j(\text{there})$.}
\begin{equation}\label{voS}
\begin{aligned}
     & V_{j,m,\bar{m}}(z,\bar{z})=\int d^2x \ x^{j+m}\bar{x}^{j+\bar{m}}\Phi_{-1-j}(x,\bar{x}|z,\bar{z})\\
     & \Phi_{j}(x,\bar{x}|z,\bar{z})=-\frac{2j+1}{\pi}\(|\hat{\Gamma}(z,\bar{z})-x|^2e^{\hat{\phi}(z,\bar{z})}+e^{-\hat{\phi}(z,\bar{z})} \)^{2j}
\end{aligned}
\end{equation}
The function $\Phi_{j}(x,\bar{x})$ satisfies the massive scalar Laplace equation in the target space
\begin{equation}\label{EOMPhi}
\begin{aligned}
     &(\nabla^2-M^2) \Phi_{j}(x,\bar{x}|z,\bar{z})=0, ~~ M^2l_{AdS}^2=(2j+2)(2j)
\end{aligned}
\end{equation}
with the following boundary condition
\begin{equation}\label{normPhi}
    \lim_{\hat{\phi} \to \infty } e^{\hat{\phi}(2j+2)} \Phi_{j}(x,\bar{x}|z,\bar{z})=\delta(\hat{\Gamma}-x)\delta(\hat{\bar{\Gamma}}-\bar{x})
\end{equation}
These properties dictate $\Phi_j$ represent a field of mass M \cite{Giveon:1998ns, Kutasov:1999xu} (see Appendix \ref{AppA} for important details on proper normalization of vertex operators).

For the purpose of this work will not need to look into more general vertex operators, see \cite{Teschner:1997ft, Teschner:1999ug, Giveon:1998ns, Kutasov:1999xu, Maldacena:2000hw, Maldacena:2000kv, Maldacena:2001km, Ribault:2005wp, Dei:2021yom, Dei:2021xgh} and \cite{Gaberdiel:2018rqv, Eberhardt:2018ouy, Eberhardt:2019ywk, Dei:2020zui}  for more details.

\subsection{Winding condensate description}\label{2.2}

Thermal AdS$_3$ has a dual description obtained by deformation of a free theory by a  winding condensate on the spatial circle \cite{Jafferis:2021ywg, Halder:2022ykw}
\begin{subequations}\label{winding2}
\begin{empheq}{align}
  & S_{\text{TAdS}}=\frac{1}{2 \pi}\int 2 d^2 \sigma (\partial \varphi \bar{\partial} \varphi+\frac{1}{4b'}\varphi R +\beta \bar{\partial} \gamma +\bar{\beta} \partial \bar{\gamma} +\frac{\pi\mu}{2b'^2} (V^{+}+ V^{-}))\\ 
  & V^{\pm}:=\ e^{\pm \frac{k}{4}(\gamma+\bar{\gamma})} \ e^{\mp  (\int_{(0,0)}^{(z,\bar{z})}\beta dz'+ \int_{(0,0)}^{(z,\bar{z})}\bar{\beta} d\bar{z}')} \ e^{b'\varphi},~~ \mu:=2b'^4 \(\frac{\mu'\gamma(b'^2)}{\pi }\)^{1/2},\\
  & b':=\frac{1}{b''}, ~~ \pi \mu'\gamma(b'^2):=(\pi \mu''\gamma(b''^2))^{1/(b''^2)}, \quad  b''=\frac{1}{\sqrt{k-2}}, ~~ \mu''=\frac{b''^2}{\pi^2}
\end{empheq}
\end{subequations}
The geometric meaning of the fields becomes apparent in large $k$ limit: $\gamma= \hat{\xi}+i \hat{\theta}, ~~ \varphi=-\sqrt{k}\hat{r}$ ($\beta$ is an auxiliary field that keeps track of the amount of winding in space circle $\hat{\theta}$), where $\hat{r}, \hat{\xi}, \hat{\theta} $ are standard global co-ordinates. At leading order in large $k$ the metric and the  Kalb-Ramond field are given by
\begin{equation}\label{TAdSfields}
	\begin{aligned}
		& ds^2=l_{AdS}^2(\sinh^2(\hat{r})d\hat{\theta}^2+\cosh^2(\hat{r})d\hat{\xi}^2+d\hat{r}^2)\\
		& B= i l_{AdS}^2 \sinh^2 (\hat{r})  \ (d\hat{\theta} \otimes d \hat{\xi}-d\hat{\xi} \otimes d \hat{\theta})\\
	\end{aligned}
\end{equation}
Putting this background at temperature $T$, as seen by the spacetime CFT$_2$ living on the boundary of thermal AdS$_3$, amounts to the following identification of coordinates 
\begin{equation}\label{TAdSCoset}
	\begin{aligned}
            & \hat{\xi} \sim \hat{\xi}  + (T l_{AdS})^{-1}, \quad \hat{\theta} \sim \hat{\theta} + 2\pi
	\end{aligned}
\end{equation}
Vertex operators of relevance  are
\begin{equation}\label{voW}
	\begin{aligned}
		\tilde{V}_{\alpha,a,\bar{a}}=\frac{1}{\pi} \ e^{a \gamma+\bar{a} \gamma} \ e^{2\alpha \varphi}, ~~ 
	\end{aligned}
\end{equation}
They are dual to the vertex operators in (\ref{voS})
\begin{equation}
    \tilde{V}_{\alpha,a,\bar{a}} \leftrightarrow V_{j,m,\bar{m}}, ~~\alpha=-\frac{j}{2b},~~  a=m-\frac{k\nu}{4}, ~~ \bar{a}=\bar{m}-\frac{k\nu}{4}
\end{equation}
These vertex operators carry no winding in either the space or  time circle. 

The focus of this paper is string theory around the Euclidean BTZ black hole. It can be described by the background given in (\ref{TAdSfields}) with a different identification\footnote{Note that trying to convert this identification to the one in (\ref{TAdSCoset}) while keeping the background metric and $B$ field fixed through field redefinition is a non-trivial exercise because in the process one generates a non-zero $B$ field %
on the boundary torus. }
\begin{equation}\label{BTZCoset}
	\begin{aligned}
            & \hat{\xi} \sim \hat{\xi}  + 4\pi^2 (T l_{AdS}), \quad \hat{\theta} \sim \hat{\theta} + 2\pi
	\end{aligned}
\end{equation}
In EBTZ background the winding around the temporal circle $\hat{\xi}$ is not conserved. This is not apparent in this presentation. This feature becomes visible in an equivalent description of EBTZ at temperature $T$  obtained by exchanging the roles of $\hat{\xi}, \hat{\theta}$ and switching the holomorphic and anti-holomorphic coordinates on the worldsheet (needed to keep the same $B$ field)
\begin{equation}\label{EBTZfields}
	\begin{aligned}
		& ds^2=l_{AdS}^2(\cosh^2(\hat{r})d\hat{\theta}^2+\sinh^2(\hat{r})d\hat{\xi}^2+d\hat{r}^2)\\
		& B= -i l_{AdS}^2 \sinh^2 (\hat{r})  \ (d\hat{\theta} \otimes d \hat{\xi}-d\hat{\xi} \otimes d\hat{\theta})\\
   & ~~~~\hat{\theta} \sim \hat{\theta}  + 4\pi^2 (T l_{AdS}), \quad \hat{\xi} \sim \hat{\xi} + 2\pi
	\end{aligned}
\end{equation}
The worldsheet theory in this background is given by\footnote{This is obtained from (\ref{winding2}) by the replacement $\gamma \to i \bar{\hat{\gamma}}, \beta \to -i \bar{\hat{\beta}}, z \to \bar{z}$ along with their complex conjugates. While presenting (\ref{EBTZfields}) we dropped hats on variables. }
\begin{subequations}\label{windingBTZ}
\begin{empheq}{align}
  & S_{\text{EBTZ}}=\frac{1}{2 \pi}\int 2 d^2 \sigma (\partial \varphi \bar{\partial} \varphi+\frac{1}{4b'}\varphi R +\beta \bar{\partial} \gamma +\bar{\beta} \partial \bar{\gamma} +\frac{\pi\mu}{2b'^2} (W^{+}+ W^{-}))\\ 
  & ~~~~~~~~~~~~~~~~~~~~~ W^{\pm}=\ e^{\pm i \frac{k}{4}(\gamma-\bar{\gamma})} \ e^{\pm i   (\int_{(0,0)}^{(z,\bar{z})}\beta dz'- \int_{(0,0)}^{(z,\bar{z})}\bar{\beta} d\bar{z}')} \ e^{b' \varphi}%
\end{empheq}
\end{subequations}
The coupling constant $\mu$ is still given by its expression in (\ref{winding2}).
Geometrically we still identify $\gamma=\hat{\xi}+i\hat{\theta}, \hat{\theta} \sim \hat{\theta}  + 4\pi^2 (T l_{AdS}), \hat{\xi} \sim \hat{\xi} + 2\pi$ as a result $W^\pm$ above carries non-trivial winding on the temporal circle $\hat{\xi}$  as opposed to $V^\pm$ which carries non-zero winding on the spatial circle $\hat{\theta}$.

\section{Thermal free energy}\label{3}

The goal of this section is to calculate the leading order in $g_s$ partition function of string theory in the EBTZ background staying entirely within the framework of on-shell string theory.  

We will proceed to evaluate the partition function of the worldsheet CFT on unit sphere. This includes AdS$_3$, compact manifold M and ghosts, such that the total central charge vanishes. This ensures that the partition function of the worldsheet CFT is well-defined and independent of the radius of the sphere \cite{Zamolodchikov:2001dz}.
\begin{equation}
    \begin{aligned}
&  Z_{CFT}((T l_{AdS})^{-1}, l_{AdS}/l_s)= Z^M Z^{AdS_3}  Z_c
    \end{aligned}
\end{equation}
The sphere partition function of the worldsheet CFT with target space M is given by $Z^M$.  We will not discuss the evaluation of $Z^M$, except the assumption that the integration of compact zero modes on $M$ is normalized to give a factor of volume $V_M$ in string units, and contribution of non-zero modes are sub-leading in large $k$ limit, more precisely
\begin{equation}
	  Z^{M}=Z_{0}^{M}  C_{S^2}^{M}, ~~ Z_{0}^{M}=V_M, ~~C_{S^2}^{M}= 1+\mathcal{O}\(\frac{1}{k}\)
\end{equation}
The AdS$_3$ part of the partition function is defined with the measure as in \cite{Halder:2022ykw} - the normalization of the measure on Liouville-like field $\varphi$ is the same as in \cite{Zamolodchikov:1995aa}, and the measure  of the non-zero modes of the $\beta \gamma$ system is same as 
in \cite{Polchinski:v1}. 
To understand the zero mode structure of the  $\beta\gamma$ system we note that the free $\beta\gamma$ system is a type of `toric' sigma model whose Hilbert is analyzed in section 5 of  \cite{Frenkel:2005ku}. After we integrate out the radial constant mode we will be left with the free $\beta\gamma$ system (see next subsection for explanation).  The zero mode contribution from translation in $\hat{\xi}, \hat{\theta}$ gives
\begin{equation}\label{ZeroModes}
    Z^{AdS_3}_0= 8\pi^3 (T l_{AdS})  
\end{equation}
The non-zero mode $\gamma', \bar{\gamma'}$ contributes as
\begin{equation}\label{nonzeromodes}
    \begin{aligned}
& C_{S^2}^{AdS_3}=\int d\varphi d\beta' d\gamma' d\bar{\beta'} d\bar{\gamma'}e^{-S_{\text{EBTZ}}}
    \end{aligned}
\end{equation}
These two factors $ C_{S^2}^{AdS_3}, Z^{AdS_3}_0$ together give $Z^{AdS_3}$. The remaining $k,\beta$ independent factor $Z_c$, taking care of the contribution of $b,c$ ghosts (omitting the ghost zero modes).

The same factors  $Z^{M},Z_c$ appear in the calculation of the effective three-dimensional Newton's constant. Therefore when the entropy of the BTZ black hole is expressed in terms of the three-dimensional Newton's constant these factors disappear. For this reason we will not be determine them in this paper explicitly.

\subsection{Non-zero modes on AdS$_3$}\label{3.1}
\subsubsection{Radial constant mode}
First we focus on the radial constant mode $\varphi_c$ defined by 
\begin{equation}
	\varphi(z,\bar{z})=\varphi_c+\varphi'(z,\bar{z})
\end{equation}
where $\varphi_c$ denotes the kernel of the scalar Laplacian and $\varphi'(z,\bar{z})$ is the space of functions orthogonal to the kernel, using the following identity \cite{PhysRevLett.66.2051, Dorn:1994xn, Zamolodchikov:1995aa, Teschner:2001rv}
\begin{equation}
	\begin{aligned}
	\int_{-\infty}^{+\infty}d\varphi_c \ e^{2a\varphi_c-\alpha e^{2b\varphi_c}}=\frac{1}{2b}\Gamma\(\frac{a}{b}\)\alpha^{-\frac{a}{b}}
	\end{aligned}
\end{equation}
We can integrate out $\varphi_c$ exactly to get
\begin{equation}\label{resc}
	\begin{aligned}
		C_{S^2}^{AdS_3}
		=\frac{2\pi}{b'}\Gamma\(-2s'\) \( \frac{\mu}{2b'^2}\)^{2s'} \int d\varphi' d\beta' d\gamma' d\bar{\beta'} d\bar{\gamma'}e^{-S_{\text{EBTZ}}|_{\mu=0}} (\int  d^2 z \ (W^{+}(z,\bar{z})+ W^{-}(z,\bar{z})))^{2s'}
	\end{aligned}
\end{equation}
 Here we have defined
 \begin{equation}
      s'=\frac{1}{b'^2}
 \end{equation}

\subsubsection{Wick contractions}

After integrating out the radial constant mode the leftover path integration is very simple due to winding and momentum conservation - see \cite{Halder:2022ykw} for more details. Essentially only a very specific expression of the winding condensate contributes to the path integral when  $s'$ is a positive integer.\footnote{We remind the reader that $$k=2\(1+\frac{1}{s}\)$$ Therefore the calculation done for the residue is valid for small $k$. This is no surprise given that the winding condensate description is best suited for calculation in this range. This in particular implies $k$ is not an integer.  This is compatible even with compact internal manifolds, for example arising from the compactification of M theory on a Calabi-Yau fourfold singularity (see for instance \cite{Balthazar:2021xeh}). } For this special case, we perform the Wick contractions on (\ref{resc}), to write \footnote{The skeptical reader might ask for the justification of not including a factor corresponding to the partition function of the linear dilaton and free $\beta\gamma$ (and complex conjugate) system at this step. The contribution of the non-zero modes of $\beta\gamma$ system is independent of $k, \beta$, so it is assumed to be absorbed in $Z_c$. For the linear dilaton note that the curvature coupling on unit radius is only through the zero mode as a result the contribution is again independent of $k$. }
\begin{equation}\label{res}
	\begin{aligned}
		C_{S^2}^{AdS_3}
		=&\frac{2\pi}{b'}\Gamma(-2s')  \( \frac{-\mu}{2b'^2}\)^{2s'}\frac{\Gamma(2s'+1)}{\Gamma(s'+1)^2}\int\prod_{i=1}^{s'}  d^2z_i \int \prod_{i'=1}^{s'}  d^2z'_{i'} \\
		&~~~~~~~~~~~~~~~~~~~~~~\prod_{i'<j'}   \[ (z_{i'j'})  (\bar{z}_{i'j'})\] \ \prod_{i<j}   \[ (z_{ij})  (\bar{z}_{ij})\] ~~ \prod_{i,j'}   \[ (z_{ij'})^{1-k}  (\bar{z}_{ij'})^{1-k}\]
	\end{aligned}
\end{equation}
Here $z_{ij}=z_i-z_j,z_{ij'}=z_i-z'_{j'},z_{i'j'}=z'_{i'}-z'_{j'} $.
Before going forward we notice that if we change variables as
\begin{equation}
	z_i \to \lambda z_i, ~~ \bar{z}_i \to \lambda \bar{z}_i, ~~ z'_i \to \lambda z'_i, ~~ \bar{z}'_i \to \lambda z'_i, ~~ 
\end{equation}
Then all the factors of $\lambda$ cancel out since $s'(k-2)=1$
\begin{equation}
	\lambda^{2s'\times 2 +s'(s'-1)/2 \times 4+2(1-k)s'^2}=\lambda^{4s'+2s'(s'-1)-2s'^2(k-1)}
\end{equation}
Thus we have effective scale invariance even after integrating out the radial constant mode. 

We single out $z_{a}, a=1,2, z'_{a'}, a=1$ (this process works only for $s'=2,3,4,..$ - we  focus on this particular case) to write the above expression as
\begin{equation}
	\begin{aligned}
		C_{S^2}^{AdS_3}
		=&\frac{2\pi}{b'}\Gamma(-2s')  \( \frac{-\mu}{2b'^2}\)^{2s'}\frac{\Gamma(2s'+1)}{\Gamma(s'+1)^2}\int\prod_{a=1}^{2}  d^2 z_{a} \prod_{I=3}^{s'}  d^2z_I \int\prod_{a'=1}^{}  d^2z'_{a'} \int \prod_{I'=2}^{s'}  d^2z'_{I'}\\ &~~~~~~~~~~~~~~~~~~~~~~~|z_{12}|^2 |z_{11'}|^{2(1-k)}|z_{21'}|^{2(1-k)}~~\prod_{a} \[ (\prod_I |z_{aI}|^2)(\prod_{I'} |z_{aI'}|^{2(1-k)})\] \\
		&\prod_{a'} \[ (\prod_I |z_{a'I}|^{2(1-k)})(\prod_{I'} |z_{a'I'}|^{2})\] ~~~~~~\prod_{I'<J'}   \[ |z_{I'J'}|^2 \] \ \prod_{I<J}   \[ |z_{IJ}|^2  \] ~~ \prod_{I,J'}   \[ |z_{IJ'}|^{2(1-k)} \]
	\end{aligned}
\end{equation}
We make the following transformations (and their conjugates) in the given order below
\begin{equation}
	\begin{aligned}
		& z_I \to z_I+z_1, z'_{I'}\to z'_{I'}+z_1 \\
		& z_I \to z_{21} z_I, z'_{I'}\to z_{21}z'_{I'}\\
	\end{aligned}
\end{equation}
Due to the scale invariance discussed above
\begin{equation}
	\begin{aligned}
		C_{S^2}^{AdS_3}
		=&\frac{2\pi}{b'}\Gamma(-2s')  \( \frac{-\mu}{2b'^2}\)^{2s'}\frac{\Gamma(2s'+1)}{\Gamma(s'+1)^2}
  \\& \int\prod_{a=1}^{2}  d^2 z_{a}  \int\prod_{a'=1}^{}  d^2z'_{a'} \prod_{I=3}^{s'}  d^2z_I \int \prod_{I'=2}^{s'}  d^2z'_{I'} \frac{|z_{12}|^2 |z_{11'}|^{2(1-k)}|z_{21'}|^{2(1-k)}}{|z_{12}|^{6+2+2(1-k)+2(1-k)}}~~\\ & \[ (\prod_I |z_{I}|^2|z_{I}-1|^2|z_{I}-x|^{2(1-k)})(\prod_{I'} |z'_{I'}|^{2(1-k)}|z'_{I'}-1|^{2(1-k)} |z'_{I'}-x|^{2})\] \\
		& ~~~~~~\prod_{I'<J'}   \[ |z_{I'J'}|^2 \] \ \prod_{I<J}   \[ |z_{IJ}|^2  \] ~~ \prod_{I,J'}   \[ |z_{IJ'}|^{2(1-k)} \]
	\end{aligned}
\end{equation}

Here we have defined $x= \frac{z_{1'1}}{z_{21}}$.
Next we do the following change of co-ordinates for all $I, i'$
\begin{equation}
	 z_I =  \frac{\tilde{z}_I x}{\tilde{z}_I+x-1}, z'_{I'}= \frac{\tilde{z}'_{I'} x}{\tilde{z}'_{I'}+x-1}, ~~  
\end{equation}
This transformation has a non-trivial Jacobian associated with it
\begin{equation}
dz_i=\frac{x(x-1)}{(z_i+x-1)^2} d\tilde{z}_i
\end{equation}
This transformation has the following properties
\begin{equation}
	\begin{aligned}
		& z_I-1=(\tilde{z}_I-1)\frac{(x-1)}{\tilde{z}_I+x-1}, ~~ z_I-x=\frac{-x(x-1)}{\tilde{z}_I+x-1}\\
		& z_I-z_J=(\tilde{z}_I-\tilde{z}_J)\frac{x(x-1)}{(\tilde{z}_I+x-1)(\tilde{z}_J+x-1)}\\
		& z_I-z_{I'}=(\tilde{z}_I-\tilde{z}_{I'})\frac{x(x-1)}{(\tilde{z}_I+x-1)(\tilde{z}_{I'}+x-1)}
	\end{aligned}
\end{equation}
The change of variables becomes simple because the factor 
\begin{equation}
	\lambda=\frac{1}{z_I+x-1}
\end{equation}
for a given $I$ drops out of the integration over $z_I$
\begin{equation}
	\lambda^{4+2 \times 2+2(1-k)+2(s'-3)+2(1-k)(s'-1) }=\lambda^{4+6-2k+2s'-6+2s'(1-k)-2(1-k)}=\lambda^{2+2s'(2-k))}=1
\end{equation}
Similarly for each $I'$ integration this factor drops out.\footnote{Essentially due to \begin{equation}
	\lambda^{4+2(1-k) \times 2+2+2(s'-2)+2(1-k)(s'-2) }=\lambda^{4+4(1-k)+2+2s'-4+2s'(1-k)-4(1-k) }=1
\end{equation}} 
As a result the integrations over $z_I, z'_{I'}$ does not involve any factor of $z_a,z'_{a'}$. We can present the result of the manipulation as
\begin{equation}
	C_{S^2}^{AdS_3}=z_{div}\  z
\end{equation}
where
\begin{equation}
	\begin{aligned}
			z_{div}=&\int\prod_{a=1}^{2}  d^2 z_{a}  \int  d^2z'_{1'} \frac{|z_{12}|^2 |z_{11'}|^{2(1-k)}|z_{21'}|^{2(1-k)}}{|z_{12}|^{6+2+2(1-k)+2(1-k)}}\frac{1}{|x|^{2(2-k)}|1-x|^{2(2-k)}}\\
			=&\int\prod_{a=1}^{2}  d^2 z_{a}  \int  d^2z'_{1'} \frac{1}{|z_{12}|^2 |z_{11'}|^{2}|z_{21'}|^2} 
	\end{aligned}
\end{equation}
and 
\begin{equation}
	\begin{aligned}
			z=&\frac{2\pi}{b'}\Gamma(-2s')  \( \frac{-\mu}{2b'^2}\)^{2s'}\frac{\Gamma(2s'+1)}{\Gamma(s'+1)^2} \prod_{I=3}^{s'}  d^2z_I \int \prod_{I'=2}^{s'}  d^2z'_{I'}~~\\ & \[ (\prod_I |z_{I}|^2|z_{I}-1|^2)(\prod_{I'} |z'_{I'}|^{2(1-k)}|z'_{I'}-1|^{2(1-k)} )\] \\
		& ~~~~~~\prod_{I'<J'}   \[ |z_{I'J'}|^2 \] \ \prod_{I<J}   \[ |z_{IJ}|^2  \] ~~ \prod_{I,J'}   \[ |z_{IJ'}|^{2(1-k)} \]
	\end{aligned}
\end{equation}
In the next two subsections, we will show show how $z_{div}$ gets cancelled against the volume of unfixed gauge group and simplify the expression of $z$ in terms of Liouville theory.

\subsection{The divergent factor}\label{3.2}

The divergent factor $z_{div}$ can be identified with the volume of the unfixed residual gauge group  PSL(2,C) (after fixing the conformal gauge). To show this we follow the footsteps of \cite{Erbin:2019uiz} and write zero temperature three-point functions in two different ways \footnote{The actual on-shell vertex operators in the BRST cohomology of string theory are $(1,1)$ on the worldsheet made up of $\Phi_{j}$ multiplied with an operator on the internal manifold M. Once this is taken into account the final conclusion made through arguments presented here remains unchanged. }
\begin{equation}\label{3pti}
    C_{S^2} \langle \Phi_{j_1}(0|0)\Phi_{j_2}(1|1)\Phi_{j_3}(\infty|\infty)\rangle = \frac{C_{S^2}}{z_{PSL(2,C)}} \int  d^2z_{1}d^2z_{2}d^2z_{3} \langle \Phi_{j_1}(0|z_1,\bar{z}_1)\Phi_{j_2}(1|z_2,\bar{z}_2)\Phi_{j_3}(\infty|z_3,\bar{z}_3)\rangle
\end{equation}
Here $C_{S^2}$ is the normalization factor for the sphere diagram at zero temperature given in (\ref{cs20}). 
The expectation value on both sides is taken in AdS$_3$ sigma model, i.e., with the measure (note that the zero mode path integration is included)
\begin{equation}
  \langle \dots \rangle=  \frac{ \int d\varphi d\beta d\gamma d\bar{\beta} d\bar{\gamma}e^{-S_{\text{EBTZ}}} \ \dots}{\int d\varphi d\beta' d\gamma' d\bar{\beta}' d\bar{\gamma}'e^{-S_{\text{EBTZ}}}}
\end{equation}
On the LHS of (\ref{3pti}) we fixed the position of three operators, thus we have inserted three factors of $c,\bar{c}$ ghosts at those locations in the complete worldsheet path integral (see section 5.3 of \cite{Polchinski:v1} for more explanation). The three-point function of ghosts cancels the $z_i,\bar{z}_i$ dependence of the matter three-point function just as in flat space and we are left with the three-point function of the matter vertex operators at three fixed points with the conformal factor involving $z_i,\bar{z}_i$ stripped off. On the RHS of (\ref{3pti}) we have not fixed the location of any vertex operator as a result there is no insertion of the ghost operator (and we are supposed to ignore the ghost zero modes in RHS\footnote{Another equivalent way of saying it is as follows: we could include the ghost zero modes in RHS of (\ref{3pti}) and then the RHS of (\ref{sl2c}) includes ghost zero mode contribution contribution. In this process we also include ghost zero modes in RHS of (\ref{nonzeromodes}) and in effect the final result is unchanged. }), however in this case we have divided by the explicit factor $z_{SL(2,C)}$ of the volume of the residual gauge group.
This gives
\begin{equation}\label{sl2c}
	\begin{aligned}
			z_{PSL(2,C)}
			=z_{div}
	\end{aligned}
\end{equation}
This is similar to the cancellation discussed in the context of Liouville theory in \cite{Mahajan:2021nsd, Giribet:2007uh}.

\subsection{The finite factor}\label{3.3}

Now we turn to simplify the finite part in terms of Liouville theory correlator. Note that in effect the finite factor is obtained by fixing location of three winding operator insertions after radial zero mode integration (just like ordinary string theoretic three point funtion, a careful consideration including $b,c$ ghosts shows in effect the ghost zero modes are taken into account). 
\begin{equation}\label{ff}
	\begin{aligned}
			z=&\frac{2\pi}{b'}\Gamma(-2s')  \( \frac{-\mu}{2b'^2}\)^{2s'}\frac{\Gamma(2s'+1)}{\Gamma(s'+1)^2} \prod_{I=3}^{s'}  d^2z_I \int \prod_{I'=2}^{s'}  d^2z'_{I'}~~\\ & \[ (\prod_I |z_{I}|^{2}|z_{I}-1|^2)(\prod_{I'} |z'_{I'}|^{2(1-k)}|z'_{I'}-1|^{2(1-k)} )\] \\
		& ~~~~~~\prod_{I'<J'}   \[ |z_{I'J'}|^2 \] \ \prod_{I<J}   \[ |z_{IJ}|^2  \] ~~ \prod_{I,J'}   \[ |z_{IJ'}|^{2(1-k)} \]
	\end{aligned}
\end{equation}
We  integrate out $z'_{I'}$ using  the integral identity in appendix B of \cite{Halder:2022ykw} with
\begin{equation}
\begin{aligned}
   & n = s'-1, \quad m = 0
   \\&  t_j = z_j,\,\, j=1,...,s'-2,\quad t_{s'-1} = 0,\, t_{s'} = 1 
   \\ & p_j = 1-k, ~~ p_{s'-1}=(1-k), ~~p_{s'}=1-k
\end{aligned}
\end{equation}
We remind the reader that the expression for the finite factor here is only valid when $s'=2,3,4... $ (for $s'=1$ we do not have enough winding operators to fix), resulting in the following formula for the residue 
\begin{equation}\label{res1st}
	\begin{aligned}
	& \quad	\textrm{Res}_{s = 2s'\rightarrow 2\mathbb{Z}} \ \ z  \\
	& = \frac{2\pi}{b'} \( \frac{-\mu}{2b'^2}\)^{2s'}\frac{1}{\Gamma(s'+1)²} \frac{\pi^{s'-1} \Gamma(s')\gamma(2-k)^{s'}}{\gamma(s'(2-k)) }   \int \prod_{I=3}^{s'} d^2z_{I} \prod_{I<J} |z_{IJ}|^{4 (2-k)} \prod_{I} |z_{I}|^{4(2-k)} |1-z_{I}|^{4(2-k)} \\
	& =\frac{2\pi}{b'} \( \frac{-\mu}{2b'^2}\)^{2s'}\frac{1}{(s'-1)(s')^2} \frac{\pi^{s'-1} \Gamma(s')\gamma(2-k)^{s'}}{\gamma(s'(2-k)) }   \Gamma(s'-1)     (- \mu')^{2-s'} \textrm{Res}_{\sum_i \alpha_i= Q' -(s'-2)b'} \,C_{(b',\mu')}(b',b',b') \\
	& =\frac{2\pi}{b'} \( \frac{-\mu}{2b'^2}\)^{2s'}\frac{1}{(s'-1)(s')^2} \frac{\pi^{s'-1} \gamma(2-k)^{s'}}{\gamma(s'(2-k)) }       (- \mu')^{2-s'} \textrm{Res}_{\sum_i \alpha_i= Q' -(s'-2)b'} \,C_{(b',\mu')}(b',b',b')
	\end{aligned}
\end{equation}
We have written down the residue using Liouville theory three-point function  (see appendix \ref{AppA} for conventions)
\begin{equation}
	C_{(b',\mu')}(\alpha_1 , \alpha_2 , \alpha_3) = \left[\pi \mu' \gamma(b'^2) b'^{2-2b'^2}\right]^{\frac{Q'-\sum_k\alpha_k}{b'}} \frac{ \Upsilon_{b'}'(0) \prod_{k=1}^{3} \Upsilon_{b'}(2\alpha_k) }{\Upsilon_{b'}(\sum_k \alpha_k - Q') \prod_{k=1}^3 \Upsilon_{b'}(\sum_k \alpha_k -2\alpha_k)}
\end{equation}
The $\Upsilon_{b'}(x)$ functions have zeros at 
\begin{equation}
	x = -m b'-nb'^{-1}, \quad x = Q'+ m b' + nb'^{-1}, \quad m,n = 0 ,1,2,...
\end{equation}
\begin{equation}
\begin{aligned}
	& \textrm{Res}_{\sum_i \alpha_i= Q' -(s'-2)b'} \,C(\alpha_1,\alpha_2,\alpha_3)   = b'\, \textrm{Res}_{s' =2 +\frac{Q' -\sum_i \alpha_i }{b'}} \,C(\alpha_1,\alpha_2,\alpha_3) \\
	 & = b'\, \textrm{Res}_{s' =2 +\frac{Q' -\sum_i \alpha_i }{b'}} \,\left[\pi \mu' \gamma(b'^2) b'^{2-2b'^2}\right]^{\frac{Q-\sum_i\alpha_i}{b'}} \frac{ \Upsilon_{b'}'(0)  \Upsilon_{b'}(2b')^3 }{\Upsilon_{b'}(- (s'-2)b')  \Upsilon_{b'}(b')^3}
\end{aligned}
\end{equation}
Here we get a pole when $s'=2,3,4,..$ from $\Upsilon_{b'}(- (s'-2)b')$. As discussed above there are precisely the values of $s'$ where we also get a pole in $z$. 
We generalize this result for all values of $s'$ as follows  \footnote{We used
\begin{equation}
	 \textrm{Res}_{2s' \rightarrow 2\mathbb{Z}} z = 2\textrm{Res}_{s' \rightarrow \mathbb{Z}} z 
\end{equation}} 
\begin{equation}\label{pf2}
\begin{aligned}
		z  &= (-1)^{2 l s'} \( \frac{-\mu}{2b'^2}\)^{2s'}\frac{\pi }{(s'-1)(s')^2} \frac{\pi^{s'-1} \gamma(2-k)^{s'}}{\gamma(s'(2-k)) } (- \mu')^{2-s'} \\& ~~~~~~~~~~~~~~~~~~~~~~~~~~~~~~~~~~~~~~~~~~~~~ \left[\pi \mu' \gamma(b'^2) b'^{2-2b'^2}\right]^{\frac{Q-\sum_i\alpha_i}{b'}} \frac{ \Upsilon_{b'}'(0)  \Upsilon_{b'}(2b')^3 }{\Upsilon_{b'}(- (s'-2)b')  \Upsilon_{b'}(b')^3}\\
  & = (-1)^{2 l s'} \( \frac{-\mu}{2b'^2}\)^{2s'}\frac{\pi}{(s'-1)(s')^2} \frac{\pi^{s'-1} \gamma(2-k)^{s'}}{\gamma(s'(2-k)) } (- \mu')^{2-s'}\\& ~~~~~~~~~~~~~~~~~~~~~~~~~~~~~~~~~~~~~~~~~~~~~~~~~~  \left[\pi \mu' \gamma(b'^2) b'^{2-2b'^2}\right]^{s'-2} \frac{ \Upsilon_{b'}'(0)  \Upsilon_{b'}(2b')^3 }{\Upsilon_{b'}(- (s'-2)b')  \Upsilon_{b'}(b')^3}  \\
		& = (-1)^{2 l s'} (-1)^{s'}\( \frac{\mu}{2b'^2}\)^{2s'}\frac{\pi}{(s'-1)(s')^2} \frac{\pi^{2s'-3} \gamma(2-k)^{s'}}{\gamma(s'(2-k)) }\\& ~~~~~~~~~~~~~~~~~~~~~~~~~~~~~~~~~~~~~~~~~~~~~~~~~~~~~~~   \left[ \gamma(b'^2) b'^{2-2b'^2}\right]^{s'-2} \frac{ \Upsilon_{b'}'(0)  \Upsilon_{b'}(2b')^3 }{\Upsilon_{b'}(- (s'-2)b')  \Upsilon_{b'}(b')^3}
\end{aligned}
\end{equation}
Here $l$ is an unfixed integer, since we are analytically continuing from integer values of $s'$, $l$ is not determined at this stage. Later by demanding reality of the resulting formula will determine $l$.

We can actually simplify the above expression a lot owing to the special argument of $\Upsilon_{b'}$. We use the following identities \cite{Fateev:2007qn}
\begin{equation}\label{Yindentitites}
    \begin{aligned}
        \Upsilon_{b'}(Q'-x)=\Upsilon_{b'}(x), ~~ \Upsilon_{b'}(x)=\frac{\Upsilon_{b'}(x+b')}{\gamma(b'x)b^{1-2b'x}}=\frac{\Upsilon_{b'}(x+\frac{1}{b'})}{\gamma(\frac{x}{b'})\(\frac{1}{b'}\)^{1-\frac{2x}{b'}}}
    \end{aligned}
\end{equation}
First factor in (\ref{pf2}) simplifies to (see (\ref{Yindentitites}))
\begin{equation}
	\frac{\Upsilon_{b'}(2b')^3}{\Upsilon_{b'}(b')^3} = \left(\gamma (b'^2) b'^{1-2b'^2}\right)^3
\end{equation}
The other factor can be manipulated as follows
\begin{equation}
\begin{aligned}
	\Upsilon_{b'}(-(s'-2)b') & = \Upsilon_{b'}(-s'b'+2b') = \Upsilon_{b'}\left(2b' - \frac{1}{b'} \right) = \gamma (b'^2 -1) b'^{3 -2b'^2 } \Upsilon_{b'}\left(b'- \frac{1}{b'}\right)  \\
	& = \gamma (b'^2 -1) b'^{3 -2b'^2 }b'^{3} \gamma(-1)  \Upsilon_{b'}\left(- \frac{1}{b'}\right)
\end{aligned}
\end{equation}
The last two factors are divergent and zero respectively. A proper limiting process can be obtained by using
 $s' = (k-2)^{-1} + \epsilon$
\begin{equation}
\begin{aligned}
    &\gamma (-1) \rightarrow \gamma (-1 - \epsilon b'^2) \  = \frac{1}{b'^2 \epsilon}\\
    &\Upsilon_{b'}\left( - \frac{1}{b'}\right) \rightarrow \Upsilon_{b'}( - \frac{1}{b'}-\epsilon b'  ) = \frac{\Upsilon_{b'}(-\epsilon b'  )}{ \gamma (-b'^{-2} ) b'^{2b'^{-1} (-b'^{-1}) -1} } = -\epsilon b' \Upsilon_{b'}'(0) \frac{b'^{\frac{2}{b'^2} + 1}}{ \gamma (-\frac{1}{b'^2} )} 
\end{aligned}	
\end{equation}
So in summary, we have
\begin{equation}
\begin{aligned}
	\Upsilon_{b'}(-(s'-2)b') &  =  - \gamma (b'^2 -1) b'^{6 -2b'^2 }    \frac{1}{b'^2 }b' \Upsilon_{b'}'(0) \frac{b'^{\frac{2}{b'^2} + 1}}{ \gamma (-\frac{1}{b'^2} )}  \\
	& =  -b'^{6 - 2 b'^2+\frac{2}{b'^2} }  \gamma (b'^2 -1)     \frac{1}{ \gamma (-\frac{1}{b'^2} )}  \Upsilon_{b'}'(0)
\end{aligned}
\end{equation}
Putting it back into (\ref{pf2}) we get 
\begin{equation}\label{pf3}
\begin{aligned}
    z &=\pi (-1)^{(2 l+1) s'} \( \frac{\mu}{2b'^2}\)^{2s'}\frac{1}{(s'-1)(s')^2} \frac{\pi^{2s'-3} \gamma(2-k)^{s'}}{\gamma(s'(2-k)) } \\
    & ~~~~~~~~~~~~~~~~~~~~~~~~~~~~~~~~~~~~~~~\left[ \gamma(b'^2) b'^{2-2b'^2}\right]^{s'-2} \frac{ \left(\gamma (b'^2) b'^{1-2b'^2}\right)^3  }{ -b'^{6 - 2 b'^2+\frac{2}{b'^2} }  \gamma (b'^2 -1)     \frac{1}{ \gamma (-\frac{1}{b'^2} )} }\\
    &= \frac{(-1)^{\frac{2l}{k-2}  }}{\Gamma(-1)} \mu^{\frac{2}{k-2}} \frac{(-1)^{\frac{k}{k-2}} 4^{\frac{1}{2-k}} (k-3) (k-2)^{\frac{3 k+2}{4-2 k}} \pi ^{\frac{2}{k-2}-2} \Gamma \left(\frac{1}{2-k}\right)}{\Gamma \left(1+\frac{1}{k-2}\right)}
\end{aligned}
\end{equation}
Now we substitute the value of cosmological constant from (\ref{winding2})
\begin{equation}
    \mu=2 (k-2)^2 \pi ^{-\frac{k}{2}} \left(-\frac{\Gamma \left(\frac{1}{k-2}\right)}{\Gamma \left(\frac{1}{2-k}\right)}\right)^{\frac{k-2}{2}}
\end{equation}
Plugging it back into (\ref{pf3}), and setting $l=-1$ to get rid of the complex phase we get
\begin{equation}\label{FinitFactor}
    \begin{aligned}
     z= &  \frac{1}{\Gamma(-1)\pi^3} \(\sqrt{k-2}-\frac{1}{\sqrt{k-2}}\)
     =  & \frac{1}{\Gamma(-1)\pi ^3} \sqrt{k} \( 1-\frac{2}{k}+\mathcal{O}\left(\frac{1}{k^2}\right)\)\\
    \end{aligned}
\end{equation}
Curiously, the dependence of $z$ on $k$ is exactly given by the slope of the linear dilaton of the dual conformal field theory on the boundary of AdS \cite{Eberhardt:2021vsx}.

\subsection{Thermal entropy}\label{3.4}
The string theoretic thermal free energy of BTZ black hole gets contributions from connected worldsheets only \cite{Polchinski:1985zf}, thus it is determined by combining results from (\ref{ZeroModes}), (\ref{sl2c}), (\ref{FinitFactor}) to be
\begin{equation}
   \log Z_{BTZ}(1/(T l_{AdS}), l_{AdS}/l_s)=\frac{1}{g_s^2} \frac{Z_{CFT}(1/(T l_{AdS}), l_{AdS}/l_s) }{ z_{PSL(2,C)}}
\end{equation}
Here $g_s$ is the string coupling at the tip of the Euclidean cigar.\footnote{In the BTZ geometry given by (\ref{EBTZfields}) the angular circle remains of the finite size at the tip, whereas the Euclidean time circle shrinks. When one takes into account the perturbative $\alpha'$ corrections to it and the effect of winding tachyon in near horizon geometry, the physical meaning of the tip might be more subtle \cite{Halder:2024gwe}.  } To the leading order in large $k$ we find
\begin{equation}\label{BTZpf}
    \log Z_{BTZ}(1/(T l_{AdS}), l_{AdS}/l_s)= \frac{8Z_c}{\Gamma(-1)} \frac{V_M}{g_s^2}\frac{\sqrt{k}}{(T l_{AdS})^{-1})}=\frac{8Z_c}{\Gamma(-1)} \frac{V_M}{g_s^2 l_s} l_{AdS} (T l_{AdS})
\end{equation}
This is precisely the formula for the entropy for BTZ black hole found from target space gravity analysis \cite{Maldacena:1998bw} (up to constant factors). We remind the reader that the radius of the horizon is related to the inverse temperature measured in AdS units $(Tl_{AdS})^{-1}$ according to
\begin{eqnarray}
    \frac{r_H}{l_{AdS}}=\frac{2\pi}{(Tl_{AdS})^{-1}}
\end{eqnarray}
 The factor $\Gamma(-1)^{-1}$ is 0, and tells us that our method is insensitive to the 
boundary terms in the target space effective action. In the next section, we will carefully discuss the conceptual origin of the overall factor  $\Gamma(-1)^{-1}$ in terms of the replica trick.

\section{Stringy replica trick}\label{4}

In the previous section, we performed the calculation of the thermal entropy of the BTZ black hole using the on-shell worldsheet description and noticed that the final result is zero. In section \ref{4.1} we will show that the technical reason for the factor of zero is a brunch cut becoming transparent. As a remedy in section \ref{4.2} we will develop a precise version of the replica trick \cite{Lewkowycz:2013nqa, Faulkner:2013ana, Engelhardt:2014gca} in the language of non-integer winding condensates.

\subsection{Understanding the zero}\label{4.1}

The integral identity in appendix B of \cite{Halder:2022ykw} was used to obtain (\ref{res1st}) that generated the factor of $\gamma(s'(2-k))^{-1}$. We turn to understand this factor of zero for the simplest case of the identity\footnote{In the notation of \cite{Halder:2022ykw}, this corresponds to 
\begin{equation}
    n=1, ~ m=0, ~ t_1=0, t_2=1, p_1=a, p_2=b
\end{equation}
} 
\begin{equation}\label{G1iden}
	\begin{aligned}
		G(a,b)=\int d^2 z |z|^{2a}|1-z|^{2b}= \pi \frac{\gamma(1+a)\gamma(1+b)}{\gamma(2+a+b)}
	\end{aligned}
\end{equation}
We write $z=x+i y$, $x,y \in \mathbb{R}$ and analytically continue the integration over $y$ as follows
\begin{equation}\label{VarChange1}
	y \to iye^{-2i \epsilon}, ~~ \epsilon \to 0+
\end{equation}
This gives
\begin{equation}
	\begin{aligned}
		G(a,b)=-\frac{i}{2}& \int_{-\infty}^{+\infty} dz_+  (z_+-i \epsilon (z_+-z_-))^a(z_+-1-i \epsilon (z_+-z_-))^b \\
		& \int_{-\infty}^{+\infty} dz_- (z_-+i \epsilon (z_+-z_-))^a(z_--1+i \epsilon (z_+-z_-))^b
	\end{aligned}
\end{equation}
Here we defined $z_\pm=x\pm y$. Say $z_+ \in (-\infty,0)$, the contour of integration over $z_-$ runs below the singularity at $z_-=0,1$. Provided the integration contour can be deformed away near infinity on the lower half-plane, the contribution is zero.
A similar argument holds for $z_+ \in (1, \infty)$ (in this case we can deform the contour in the upper half plane). For $z_+ \in (0,1)$, the contour of integration over $z_-$ runs above the singularity at $z_-=0$ and below at $z_-=1$. We can deform the contour to run around $z_- \in (1, \infty)$ encircling the point at $z_- =1$. If the integration over the circle around $z_- =1$ does not contribute anything, we get
\begin{equation}\label{trans}
\begin{aligned}
     G(a,b)&=-\sin (\pi b)\int_{0}^{1} dz_+  (z_+)^a (1-z_+)^b
 \int_{1}^{\infty} dz_-  (z_-)^a (z_--1)^b\\
 &=-\sin (\pi b)\int_{1}^{\infty} dz_+  (z_+)^{-a-b-2} (z_+-1)^b
  \int_{1}^{\infty} dz_-  (z_-)^a (z_--1)^b
\end{aligned}
\end{equation}
The factor of $\sin(\pi b)$ comes from the fact that the integration around the branch cuts starting at $z_- =1$ has the opposite orientation. To go from the first to the second line we changed variables from $z_+\to 1/z_+$. It is easy to check that this gives the same answer as in (\ref{G1iden}) by using
\begin{equation}
	\int_{0}^{1} dz_+  (z_+)^a (1-z_+)^b=\frac{\Gamma(1+a)\Gamma(1+b)}{\Gamma(2+a+b)}, ~~ \Gamma(z)\Gamma(1-z)=\frac{\pi}{\sin(\pi z)}
\end{equation} 
The first factor in the second line of (\ref{trans}) shows when 
\begin{equation}
    2+a+b=0,-1,-2,... 
\end{equation}
we have a transparent branch cut that makes the integral vanish. This explains the origin of the overall factor of $\Gamma(-1)^{-1}$ we had in our calculation in the previous section. A branch cut becoming transparent is highly suggestive of the smooth limit of a replica geometry. We make this precise in the next sub-section.

\subsection{Getting rid of the zero}\label{4.2}

In this sub-section, we will propose a precise version of the replica trick on the worldsheet that is valid for all orders in $\alpha'$. In order to do that we need to first discuss operators that create a conical singularity in the target space. Given that this is a subtle topic we proceed for notational convenience by first discussing it in the thermal AdS$_3$ background made out of a spatial winding condensate, and then proceed to Euclidean BTZ.

A generic vertex operator in TAdS$_3$ carrying spatial winding $\nu$  takes the following form  (see (see \cite{Halder:2022ykw} for more details)
\begin{equation}\label{vo}
	\begin{aligned}
\tilde{V}^\nu_{\alpha,a,\bar{a}}=N^\nu_{\alpha,a,\bar{a}} \ e^{a \gamma+\bar{a} \gamma} \ e^{n(\int_{(0,0)}^{(z,\bar{z})}\beta dz'+ \int_{(0,0)}^{(z,\bar{z})}\bar{\beta} d\bar{z}')} \ e^{2\alpha \varphi}
	\end{aligned}
\end{equation}
with
\begin{equation}
	\begin{aligned}
		& \alpha:=-\frac{j}{b'},~~ 
		& a:=m-\frac{k\nu}{4}, ~~
		& n:=-\nu
	\end{aligned}
\end{equation}
We start with a vertex operator with  no spatial winding, i.e., $\nu=0$ and with angular momentum 
\begin{equation}\label{TAdSm}
    m=\bar{m}=\frac{k}{2}
\end{equation}
At this stage the radial momentum $j$ is arbitrary. 
We proceed by turning on $\nu= 1$ unit of spatial winding. The resulting operator carries no momentum in any circle.
\begin{equation}
    m'=m-\frac{k}{2}\nu=0,~~ \bar{m}'=\bar{m}-\frac{k}{2}\nu=0
\end{equation}

Then $j$ is fixed  by requiring the resulting operator to be on-shell (and decaying at the AdS boundary  $\varphi\to -\infty$) in the free theory
\begin{equation}
    j=1-\frac{k}{2}
\end{equation}
This gives us $V^+$.

The off-shell background for the replica trick with $2\pi (1-\delta)$ opening angle is obtained as follows. We start with a vertex operator carrying no winding and momentum close to \eqref{TAdSm}.  
\begin{equation}\label{fracMomemtum}
    m=\bar{m}=\frac{k}{2}(1-c \delta)
\end{equation}
Here $c$ is an order one constant. After turning on winding
\begin{equation}\label{fracWinding}
    \nu=1-\delta
\end{equation}
we require that the resulting operator conserves momentum on both circles which fixes\footnote{Here we are assuming the expression of the currents in terms of free fields remain unchanged as we introduce the conical singularity. }
\begin{equation}
    c=1
\end{equation}
Now we fix the radial momentum by requiring the operator to be $(1,1)$ to get
\begin{equation}\label{replicaAction1}
\begin{aligned} 
  & V_{\delta}^{\pm}:=\ e^{\pm \frac{k}{4}(1-\delta)(\gamma+\bar{\gamma})} \ e^{\mp (1-\delta) (\int_{(0,0)}^{(z,\bar{z})}\beta dz'+ \int_{(0,0)}^{(z,\bar{z})}\bar{\beta} d\bar{z}')} \ e^{ \(1-\frac{1}{1-\frac{3}{k}}\delta\) \sqrt{k-2} \varphi }
\end{aligned}
\end{equation}
 This operator is not on-shell in the interacting theory because it does not have integer quantized winding (so it is not even a local operator). 
Consider the theory obtained by replacing $V^\pm$  with $V_{\delta}^{\pm}$ in the action of TAdS$_3$ (\ref{winding2}). The operator $V_{\delta}^{\pm}$ has a non-zero reflection coefficient of order $\delta$. Therefore to have a smooth interior we would have to add the reflected wave as well in the action (moreover, the reflected wave would change the AdS asymptotics for $k>3$). Our proposal is that we just add $V_{\delta}^{\pm}$ to the action and do not add any reflected wave associated with it. This keeps the AdS asymptotics unchanged but creates a conical singularity in the $\hat{r},\hat{\theta}$ plane.

This discussion is easily generalized to Euclidean BTZ and we propose the replica geometry required for the evaluation of the BTZ black hole entropy is given by replacing $W^\pm$ by\footnote{Note that $W^\pm_\delta$ remains weight $(1,1)$ if we change the coordinate periodicities to  $\hat{\theta} \sim \hat{\theta}  + 4\pi^2 (T l_{AdS}), \hat{\xi} \sim \hat{\xi} + 2\pi(1-\delta)$, which corresponds to a background with asymptotic temperature $T (1+\delta)$. Then it would have been a conventional on-shell vertex operator.
}
\begin{equation}\label{replicaAction}
\begin{aligned} 
  &W_{\delta}^{\pm}=  e^{\pm i \frac{k}{4}(1-\delta)(\gamma-\bar{\gamma})} \ e^{\pm i(1-\delta)   (\int_{}^{(z,\bar{z})}\beta dz'- \int_{}^{(z,\bar{z})}\bar{\beta} d\bar{z}')} \ e^{ \(1-\frac{1}{1-\frac{3}{k}}\delta\) \sqrt{k-2} \varphi }
\end{aligned}
\end{equation}
in (\ref{windingBTZ}) (with no reflected wave added).

For the purpose of calculating the thermal entropy, we need to understand the variation of the free energy of this theory to first order in $\delta$. We need to evaluate the free energy with the insertion of winding operators $W_{\delta}^{\pm}$. However, since evaluation of entropy requires only the knowledge of the first derivative with respect to $\delta$ we can set all $W_{\delta}^{\pm}$ to $W_{\delta=0}^{\pm}$ except one. In the context of the calculation in the previous section, we fix the $W^+_{\delta \neq 0}$ to be at the origin.

After carefully taking into account the symmetry factor ($2s'$),  the regulated expression is given by (all the terms of order $\delta^2$ or higher are thrown away)
\begin{equation}
	\begin{aligned}
			z=&\frac{2\pi}{b'}\Gamma(-2s')  \( \frac{-\mu}{2b'^2}\)^{2s'}\frac{\Gamma(2s'+1)}{\Gamma(s'+1)^2} \prod_{I=3}^{s'}  d^2z_I \int \prod_{I'=2}^{s'}  d^2z'_{I'}~~\\ & \[ (\prod_I |z_{I}|^{2(1+\frac{ k}{k-3} s'\delta)}|z_{I}-1|^2)(\prod_{I'} |z'_{I'}|^{2(1-k+(2k+\frac{k}{k-3})s'\delta)}|z'_{I'}-1|^{2(1-k)} )\] \\
		& ~~~~~~\prod_{I'<J'}   \[ |z_{I'J'}|^2 \] \ \prod_{I<J}   \[ |z_{IJ}|^2  \] ~~ \prod_{I,J'}   \[ |z_{IJ'}|^{2(1-k)} \]
	\end{aligned}
\end{equation}
Now we turn to simplify this expression in terms of Liouville theory correlators.
We  integrate out $z_{I'}$ with 
\begin{equation}
\begin{aligned}
   & n = s'-1, \quad m = 0
   \\&  t_j = z_j,\,\, j=1,...,s'-2,\quad t_{s'-1} = 0,\, t_{s'} = 1 
   \\ & p_j = 1-k, ~~ p_{s'-1}=1-k+a(k)\delta, ~~p_{s'}=1-k
\end{aligned}
\end{equation}
Interestingly, we have a $k$ dependent factor in the regulator 
\begin{equation}
    a(k)=\(2+\frac{1}{k-3} \) \(\frac{k}{k-2}\)
\end{equation}

We remind the reader that the expression for the finite factor here is only valid when $s'=2,3,4... $ (for $s'=1$ we do not have enough winding operators to fix).  This gives the following formula for the residue at those values
\begin{equation}\label{manipulation}
	\begin{aligned}
	& \quad	\textrm{Res}_{s = 2s'\rightarrow 2\mathbb{Z}} \ \ z  \\
	& = \frac{2\pi}{b'} \( \frac{-\mu}{2b'^2}\)^{2s'}\frac{1}{\Gamma(s'+1)²} \frac{\pi^{s'-1} \Gamma(s')\gamma(2-k)^{s'-1}\gamma(2-k+a(k)\delta)}{\gamma(s'(2-k)+a(k)\delta) } 
 \\& ~~~~~~~~~~~~~~~~~~~~~~~~~~~~~~~~~~\int \prod_{I=3}^{s'} d^2z_{I} \prod_{I<J} |z_{IJ}|^{-4 (k-2)} \prod_{I} |z_{I}|^{-4(k-2)(1-\frac{k}{(k-3)(k-2)}\delta)} |1-z_{I}|^{-4(k-2)} \\
	& =\frac{2\pi}{b'} \( \frac{-\mu}{2b'^2}\)^{2s'}\frac{1}{(s'-1)(s')^2} \frac{\pi^{s'-1} \Gamma(s')\gamma(2-k)^{s'-1}\gamma(2-k+a(k)\delta)}{\gamma(-1+a(k)\delta) } 
 \\& ~~~\Gamma(s'-1)     (- \mu')^{2-s'} \textrm{Res}_{\sum_i \alpha_i= Q' -(s'-2)b'} \,C_{(b',\mu')}(b',b'(1-\frac{k}{(k-3)(k-2)}\delta),b'(1+\frac{k}{(k-3)(k-2)}\delta)) \\
	& =-a(k) \delta \ \frac{2\pi}{b'} \( \frac{-\mu}{2b'^2}\)^{2s'}\frac{1}{(s'-1)(s')^2} \pi^{s'-1} \gamma(2-k)^{s'}       (- \mu')^{2-s'} \textrm{Res}_{\sum_i \alpha_i= Q' -(s'-2)b'} \,C_{(b',\mu')}(b',b',b')
	\end{aligned}
\end{equation}
In the second line of (\ref{manipulation}), we have written down the residue using the Liouville theory three-point function. To go from the second to the third line of (\ref{manipulation}) we have kept only terms of leading order in $\delta \to 0$ limit. This is precisely the expression in section \ref{3.3} up to the overall factor of the regulator $-a(k)\delta$.

In this language, the entropy is obtained from
\begin{equation}\label{replicaEE}
    S_{BTZ}=\lim_{\delta\to 0}\(-\partial_{n}(\log Z_{BTZ}(n)-n\log Z_{BTZ}(1)) \bigg|_{n=1+\delta}\)
\end{equation}
We get the following exact in $l_{AdS}/l_s$ expression for the entropy of the BTZ black hole
\begin{equation}\label{finalEntropy}
\begin{aligned}
    S_{BTZ}
    =& 16 Z_c \( \frac{Z^M}{g_s^2 \sqrt{k-2}}  \) k \(1-\frac{1}{2(k-2)}\)   \ (Tl_{AdS})\\
    = & \(\frac{1}{g_s^2 } Z_c Z^M  \ Tl_{AdS} \) \  16 \sqrt{k} \( 1+\frac{1}{2k}+\mathcal{O}\left(\frac{1}{k^2}\right)\)
\end{aligned}
\end{equation}
On the other hand, from the target space gravity analysis the entropy of the BTZ black hole is given by 
\begin{equation}
    \begin{aligned}
       & S_{BTZ} = \frac{2\pi V_M}{g_{s}^2} \( 2\pi T l_{AdS} \) \(2\pi \sqrt{k}\)\( 1+\mathcal{O}\left(\frac{1}{k}\right)\)
    \end{aligned}
\end{equation}
The numerical coefficient matches precisely if the path integral is normalized according to
\begin{equation}
    Z_c=\frac{\pi^3}{2}
\end{equation}

\subsection{The area operator}\label{4.3}

The calculation performed in this note makes it clear that for the purpose of evaluating thermal entropy, we need to only go away from the on-shell background by an infinitesimal amount, in particular, this leads to the fact that entropy comes entirely from one point function of the stringy area operator
\begin{equation}\label{areaOp}
\begin{aligned}
    & A=\frac{ \mu}{b'^4} \int d^2 \sigma \  \frac{\partial}{\partial \delta} (W^+_\delta+W^-_\delta)\bigg|_{\delta=0}\\
\end{aligned}
\end{equation}
This operator is not a local operator, in the sense that it has a logarithmic branch cut attached to it. This can be seen from the following OPE
\begin{equation}
	\begin{aligned}
	&	W_\delta^+(z, \bar{z}) \tilde{V}^{0}_{\alpha,a,\bar{a}}(0,0) \sim (z)^{-\alpha b'(1-\frac{1}{1-\frac{3}{k}}\delta)-im(1-\delta)} (\bar{z})^{-\alpha b'(1-\frac{1}{1-\frac{3}{k}}\delta)+i\bar{m}(1-\delta)}(1+ \dots) \\ 
	&	W_\delta^-(z, \bar{z}) \tilde{V}^{0}_{\alpha,a,\bar{a}}(0,0) \sim (z)^{-\alpha b'(1-\frac{1}{1-\frac{3}{k}}\delta)+im(1-\delta)}(\bar{z})^{-\alpha b'(1-\frac{1}{1-\frac{3}{k}}\delta)-i\bar{m}(1-\delta)}(1+\dots)\\
	\end{aligned}
\end{equation}
Because of these branch cuts the area operator $A$ given in (\ref{areaOp}) is not a local operator and therefore evaluation of its one-point function is tricky (see a similar discussion for non-normalizable operators in \cite{Kraus:2002cb}).

\section{Summary and future directions}\label{5}

In this paper, we have calculated the thermal Bekenstein-Hawking entropy of the BTZ black hole in bosonic string theory on AdS$_3\times$M from a first principles analysis using the winding condensate description of the worldsheet theory. Here, we want to emphasize to the reader the significance of the derivation and its underlying logic. From the point of view of the worldsheet, there are two dimensionless quantities 
\begin{equation}
    \frac{l_{AdS}}{l_{s}}, ~~ T l_{AdS}
\end{equation}
The free energy is order $g_s^{-2}$ on a spherical worldsheet due to the usual contribution of the string dilaton. Integration over the angular zero modes of the worldsheet fields in the BTZ background produces a factor of $T l_{AdS}$ in free energy. At this point, the dependence on $l_{AdS}/l_s$ is completely unknown (the curious reader might demand here that we use the additional input from target space physics that entropy must be order $G_N^{-1}$, and since $G_N$ has a scale it will fix the $l_{AdS}$  dependence of the answer when $l_{AdS}/l_s$ is scaled to infinity). The central point of this paper is to avoid any such inputs from the target space physics and derive the result intrinsically. 
Naively one might expect a very simple answer for the coefficient of the linear in temperature dependence of the entropy of the BTZ blackhole,\footnote{As an observation we note that the explicit $k$ dependence in (\ref{finalEntropy}) is the same as the product of the central charge of the SL(2,R) WZW model and the maximum allowed  Liouville-like momentum of an operator in the boundary conformal field theory corresponding to the discrete representations \cite{Eberhardt:2021vsx}. } based on AdS$_3$/CFT$_2$, in terms of central charge of the dual conformal field theory (although there are potential subtleties in the present context due to long strings rendering the dual theory not a standard CFT$_2$, as the spectrum is not discrete, and our setting in the bosonic string which has a closed string tachyon, so it is not truly a valid background). However, it is not clear that the states which contribute to the Cardy formula of the boundary CFT all come from black holes as we enter the highly stringy regime.\footnote{Arbitrary higher derivative corrections to the entropy of non-rotating BTZ black holes can be put in a suggestive form of a Cardy formula, i.e, the entropy is proportional to the temperature (we are considering a non-chiral theory), however it is not clear if the coefficient of proportionality is given by the central charge of the dual CFT$_2$ in general. } Moreover, the nature of BTZ as a quotient makes it clear that the worldsheet genus zero free energy is exactly linear in T, since the temperature therefore only enters through the volume of zero modes, and there are no worldsheet instantons corresponding to maps of S$^2$ into the angular directions of the target space which is contractible to S$^1$.  In this work, we determined the exact dependence on $k$ using the following  ingredients - 
\begin{enumerate}
    \item We performed an honest path integration of the AdS$_3$ part of worldsheet CFT, however, we did not attempt to consider the path integration over the worldsheet CFT on the compact manifold $M$ in detail.
    The functional dependence of the entropy on the compact manifold M is only through an overall factor that also appears in the three-point function of vertex operators contained entirely in AdS$_3$ (such vertex operators always exist in compactifications of the type AdS$_3\times$M). Therefore, the dependence on M is absorbed in the definition of effective lower dimensional Planck constant. Keeping this in mind, it will be very interesting to compare our results against higher derivative corrections in the target space \cite{Kraus:2005vz, Kraus:2005zm, David:2007ak}. We leave this to future work.
    \item  We proposed a prescription for the replica trick based on the expectation that non-integer winding introduces a conical singularity at the origin. This procedure is uniquely fixed by the symmetries.  In summary, the calculation presented here is correct up to the leading order in small $g_s$, and we have managed to identify the $\alpha'$ exact `area' operator whose one-point function produces the leading order entropy of the BTZ black hole.  Given the explicit form of the area operator (\ref{areaOp}), we can extend the calculation of its one-point function to higher orders in $g_s$, say when the worldsheet is a torus. From the point of view of the replica trick in the bulk effective field theory, the one-loop contribution to the thermal entropy comes from two pieces -   the entanglement of the bulk fields (including gravitons) outside the horizon of the classical black hole geometry, and the first derivative of the log of the partition function of the replica geometry, having a conical singularity in the interior, with respect to (the inverse of) the opening angle of the conical singularity. Therefore it is likely that the one-point function of the area operator is related to the latter and presumably the former corresponds to the correction to the entropy obtained through the use of the orbifold method developed in \cite{Witten:2018xfj, Dabholkar:2022mxo}. It is a fascinating open problem to explore these questions further possibly keeping in mind any potential connection to logarithmic conformal field theories \cite{Creutzig:2013hma} and algebra of observables in the target space \cite{Leutheusser:2021qhd, Leutheusser:2021frk, Witten:2021unn, Chandrasekaran:2022eqq}.\footnote{Also see the prospectives in \cite{Verlinde:2020upt}.}
\end{enumerate}

Any consistent quantum theory of gravity must provide a microscopic explanation for the thermal  Bekenstein-Hawking entropy  \cite{PhysRevD.7.2333, Hawking:1975vcx}   of a black hole. 
In the context of string theory, Strominger and Vafa showed the microstates of a certain black hole at zero temperature belong to the Hilbert space of the theory living on the world volume of the constituent D-branes \cite{Strominger:1996sh} (see also Sen's work on black holes built out of elementary strings \cite{Sen:1995in, Sen:2004dp}). Later, Gokakumar and Vafa formulated a different approach to microstate counting in M theory exploiting its connection with topological strings \cite{Gopakumar:1998ii, Gopakumar:1998jq}.\footnote{For the successful evaluation of BPS entropy along these lines see \cite{HalderLin}.}  Over the years these techniques greatly developed  \cite{Kinney:2005ej, Bhattacharya:2008zy, Ooguri:2004zv, Denef:2007vg, Benini:2015eyy},  however, the reliance on supersymmetry\footnote{For more discussion on the contribution of complex thermal geometries to supersymmetric index see \cite{Benini:2015eyy, Cabo-Bizet:2018ehj, Choi:2018hmj, Iliesiu:2021are, H:2023qko} (see also \cite{Chen:2023mbc, Chen:2023lzq}).} implied that the explanation of the thermal part of the entropy remained elusive \cite{Horowitz:1996ay}. For instance, in the example of black  D3 branes, the temperature dependence of the entropy can be obtained using extensivity and conformal properties of the theory on the brane \cite{Gubser:1996de}, however, dependence on the 't Hooft coupling remains unknown. The similar scaling of entropy with temperature in D0 brane quantum mechanics is determined in  \cite{Biggs:2023sqw} based on the added assumption of the existence of certain scaling behavior of the effective theory of the `moduli' \cite{Paban:1998ea, Paban:1998qy}\footnote{In the context of disorder averaged Sachdev-Ye-Kitaev (SYK) model and Jackiw–Teitelboim (JT) gravity, the leading order thermal entropy of a black hole (further quantum corrections to thermal entropy in JT gravity is calculated in \cite{Heydeman:2020hhw, Iliesiu:2020qvm}) has been evaluated on the field theory side by Maldacena and Stanford \cite{Maldacena:2016hyu}.}. 
Conceptually, thermal physics is inherently much more complicated due to the presence of chaotic dynamics.\footnote{At strong coupling, black holes act as the maximally chaotic quantum systems \cite{Maldacena:2015waa, Halder:2019ric,  Camanho:2014apa, Chowdhury:2019kaq, Chandorkar:2021viw, Blake:2021wqj}.} 
 Nevertheless, in this paper, we have presented a \textit{well-controlled} worldsheet exact calculation for the thermal Bekenstein-Hawkin entropy of the BTZ black hole.\footnote{For a discussion of absorption coefficients in AdS$_3$/CFT$_2$ based on thermal two-point functions see \cite{Das:1996wn, Das:1996jy, Dhar:1996vu, Maldacena:1997ih}. }. A state counting picture of our calculation is still lacking. It would be fantastic to interpret the calculation in terms of the state counting in the Hilbert space of the worldsheet in angular quantization \cite{Agia:2022srj} following the proposal in \cite{Jafferis:2021ywg} for the Lorentzian meaning of time winding operators, or through the effective `mini-superspace' approximation of \cite{Dijkgraaf:1991ba, Mertens:2013zya}.

The major success of the target space version of the replica trick is to produce systematic quantum corrections to the Hubeny-Rangamani-Ryu-Takayanagi formula \cite{Ryu:2006bv, Hubeny:2007xt} for the entanglement entropy of a region on the spacetime CFT and its crucial role in the discussion of the black hole information paradox (consult \cite{Almheiri:2020cfm, Raju:2020smc} and references therein for a review). It would be very interesting to understand this generalization from the thermal entropy to entanglement entropy of a sub-region of the spacetime CFT directly on the worldsheet.\footnote{See \cite{Hartnoll:2015fca, Das:2020jhy, Das:2020xoa,  Das:2022njy, Balasubramanian:2018axm} for some of the related discussions.} Perhaps a good starting point for discussion involves understanding the area operator here in terms of metric deformation using the map of the tachyon vertex operator to the graviton vertex operator as given in \cite{Giveon:2016dxe, 
 Martinec:2020gkv, Martinec:2023zha}.

\acknowledgments

 We thank Yiming Chen for  collaboration in the initial stage of this project. %
 We thank  David Kolchmeyer, Prahar Mitra, David Tong,  Aron Wall, Xi Yin, and especially Nicholas Agia, Amr Ahmadain, Juan Maldacena and Douglas Stanford for insightful discussions and comments. IH is supported by the Harvard Quantum Initiative Fellowship. The work of DLJ and IH is supported in part by DOE grants DE-SC0007870 and DE-SC0021013.

\appendix

\section{Three point function at zero temperature}\label{AppA}

In this Appendix, we will compare the three-point function as calculated from the worldsheet with that of the gravity in target space for large values of $l_{AdS}/l_s$. The process is not completely fixed in principle because of the unknown normalization of vertex operators (in flatspace string theory, the normalization of vertex operators is unambiguously fixed by demanding the unitarity of spacetime S matrix). In AdS, there is no notion of S-matrix. One might hope to take advantage of the unitarity of the spacetime CFT living on the boundary of AdS. This procedure is tricky in the AdS$_3$ background we are talking about. This is because of the issues related to the normalization of spacetime stress tensor \cite{Bertle:2020sgd}, which in turn related to the issues of normalization of the vacuum \cite{Kim:2015gak}. We will fix the normalization of vertex operators by looking at the example of AdS$_5\times S^5$ up to an order one constant factor.

It will be easier to work in the Poincare patch in $l_{AdS}=1$ unit for this purpose:
\begin{equation}
    ds^2=d\hat{\phi}^2+e^{2\hat{\phi}}d\hat{\Gamma}d\hat{\bar{\Gamma}}=\frac{du^2+d\vec{x}^2}{u^2}
\end{equation}
Here we used the following change of variables
\begin{equation}
    u=e^{-\hat{\phi}}, ~~ d\vec{x}^2=d\hat{\Gamma}d\hat{\bar{\Gamma}}
\end{equation}
By the complete analogy with the discussion of AdS$_5\times S^5$ in \cite{Lee:1998bxa}, we demand the following effective action for a set of three  scalar fields $\phi_i$ of mass $M_i$
\begin{equation}
    S_g=\frac{1}{2}\int d^2\vec{x}du \sqrt{g}\ \(g^{\mu \nu} \partial_\mu  \phi \partial_\nu \phi +M^2 \phi^2+\sqrt{l_3} \phi_1\phi_2\phi_3\), ~~ \frac{1}{l_3}=\frac{V_M}{g_s^2 l_s}
\end{equation}
In direct quantization, the bulk-to-boundary propagator $K_{\Delta_i}(u,\vec{x};\vec{y})$ from $(u,\vec{x})$ to $\vec{y}$ satisfies 
\begin{equation}\label{EOMBbp}
\begin{aligned}
     &(\nabla^2-M_i^2) K_{\Delta_i}(u,\vec{x};\vec{y})=0, ~~ \Delta_i=1+\sqrt{1+M_i^2}
\end{aligned}
\end{equation}
with the following boundary condition
\begin{equation}\label{normBbp}
    \lim_{u \to 0} u^{\Delta_i-2} K_{\Delta_i}(u,\vec{x};\vec{y})=\delta^{(2)}(\vec{x}-\vec{y})
\end{equation}
Comparing (\ref{EOMBbp}),(\ref{normBbp}) with (\ref{EOMPhi}),(\ref{normPhi}) (note the former is calculated in $l_{AdS}=1$ units, whereas later is calculated in $l_s=1$ unit) suggests that $\Phi_j$ represents a scalar field in the target space with the map $\Delta(j)=-2j$. To make this claim sharp we calculate the three-point function using the Witten diagram. Up to a constant factor it is given by \cite{Freedman:1998tz}
\begin{equation}\label{3ptWittenDiagram}
    C_{123}=\sqrt{l_3} \Gamma\( \frac{\Delta}{2}-1\)\prod_{i=1}^3\frac{\Gamma\(\frac{\Delta-2\Delta_i}{2}\)}{\Gamma\(\Delta_i-1\)}, ~~ \Delta=\sum_{i=1}^3\Delta_i
\end{equation}

Now we turn to calculate the string theoretic three-point function from the point of view of the worldsheet. We have to multiply the three-point function of $\Phi_{j_i}$ in  SL(2,R) WZW model with the normalization of path integration chosen in this paper with the sphere partition function of the worldsheet CFT. The sphere partition function is already determined in the main text for non-zero temperature. The only difference with the zero temperature calculation considered here is due to the zero mode integral in $\gamma+\bar{\gamma}$. Here the zero mode ends up giving a conservation law for the momentum of the external operators (see section 6.2 of \cite{Polchinski:v1} for more explanation), as a result, the normalization is given by
\begin{equation}\label{cs20}
    C_{S^2}=\frac{1}{g_s^2}C^{AdS_3}_{S^2}Z_0^M C^M_{S^2}\frac{Z_c}{z_{SL(2,\mathbb{C})}}\sim \frac{1}{l_{3}}
\end{equation}
Here we omitted any order one constant. 
 The three-point function in WZW model is given by \cite{Ribault:2005wp, Halder:2022ykw} (we stripped off the conformal factor involving $x,\bar{x}$)
\begin{equation}\label{sl3pt}
    \begin{aligned}
        & \langle \Phi_{j_1}(0|0)\Phi_{j_2}(1|1)\Phi_{j_3}(\infty|\infty)\rangle=\frac{1}{2\pi^3b''}\left[\frac{ \gamma(b''^2) b''^{2-2b''^2}}{\pi}\right]^{-2-\sum_k j_k} \\
        & ~~~~~~~~~~~~~~~~~~~~~~~~~~~~~~~~~~~~~~~~~~~~~\frac{ \Upsilon_{b''}'(0) \prod_{k=1}^{3} \Upsilon_{b''}(-b''(2j_k+1) }{\Upsilon_{b''}(-b''(\sum_i j_i+1)) \prod_{k=1}^3 \Upsilon_{b''}(-b''(\sum_i j_i -2j_k))}\\
    \end{aligned}
\end{equation}
For the purpose of obtaining the large $k$ limit of the formula, we use \cite{Harlow:2011ny}
\begin{equation}
\begin{aligned}
    & \Upsilon_{b}(\sigma b)=\frac{b^{-\sigma}}{\Gamma(\sigma)}F_1\sqrt{b}e^{-\frac{1}{4b^2}\log b+\frac{F_0}{b^2}+\mathcal{O}(b^2\log b)}\\
    \implies & \Upsilon'_{b}(0))=\frac{F_1}{\sqrt{b}}e^{-\frac{1}{4b^2}\log b+\frac{F_0}{b^2}+\mathcal{O}(b^2\log b)}
\end{aligned}
\end{equation}
Here $F_{0,1}$ are two constants whose value will not be important for our purpose. 
Using these formulae we get
\begin{equation}
    \begin{aligned}
      \lim_{b''\to 0} & \frac{1}{b''} \frac{ \Upsilon_{b''}'(0) \prod_{k=1}^{3} \Upsilon_{b''}(-b''(2j_k+1) }{\Upsilon_{b''}(-b''(\sum_k j_k+1)) \prod_{k=1}^3 \Upsilon_{b''}(-b''(\sum_i j_i -2j_k))}\\
       =&\Gamma\(-(\sum_i j_i+1)\) \frac{\prod_{k=1}^3 \Gamma(-(\sum_i j_i -2j_k))}{\prod_{k=1}^{3} \Gamma(-(2j_k+1))}\\
      \lim_{b''\to 0} & \gamma(b''^2) b''^{2-2b''^2}=1
    \end{aligned}
\end{equation}
This shows, the string theoretic three-point function of operators $\Phi^G_j$ defined as\footnote{There overall factor of $\sqrt{l_3}$ is obtained by keeping in mind the large central charge expansion of the spacetime CFT (two-point function in the space-time CFT is normalized to be order one). Here we introduced the factor of $\pi$ arbitrarily.}
\begin{equation}\label{mapSUGRA}
   \Phi^G_{j}=\sqrt{l_3} \ \pi^{-j_i} \ \Phi_{j} \leftrightarrow \phi \text{ with }\Delta(j)=-2j
\end{equation}
 evaluated using (\ref{sl3pt}) multiplied with (\ref{cs20}) in large $k$ limit matches precisely (up to order one factor) with the result of  Witten diagram in (\ref{3ptWittenDiagram}).

To compare our result against the one in \cite{Dei:2022pkr} (see also \cite{Eberhardt:2021vsx}), one must carefully take into account the relation between $g_s, Z^M$ and the order of the symmetric product on the spacetime CFT ($N$ in the notation of \cite{Dei:2022pkr}).\footnote{For a recent attempt see \cite{Eberhardt:2023lwd}.}  At present these relations are not known precisely because of the issues related to the normalization of spacetime stress tensor \cite{Bertle:2020sgd}, which in turn related to the issues of normalization of the vacuum \cite{Kim:2015gak} and we leave these questions to the future.

\bigskip

\providecommand{\href}[2]{#2}\begingroup\raggedright\endgroup


\begin{thebibliography}{100}

\bibitem{PhysRevD.7.2333}
J.~D. Bekenstein, {\it Black holes and entropy},  {\em Phys. Rev. D} {\bf 7}
  (Apr, 1973) 2333--2346.

\bibitem{Hawking:1975vcx}
S.~W. Hawking, {\it {Particle Creation by Black Holes}},  {\em Commun. Math.
  Phys.} {\bf 43} (1975) 199--220. [Erratum: Commun.Math.Phys. 46, 206 (1976)].

\bibitem{Ryu:2006bv}
S.~Ryu and T.~Takayanagi, {\it {Holographic derivation of entanglement entropy
  from AdS/CFT}},  {\em Phys. Rev. Lett.} {\bf 96} (2006) 181602,
  [\href{http://xxx.lanl.gov/abs/hep-th/0603001}{{\tt hep-th/0603001}}].

\bibitem{Hubeny:2007xt}
V.~E. Hubeny, M.~Rangamani, and T.~Takayanagi, {\it {A Covariant holographic
  entanglement entropy proposal}},  {\em JHEP} {\bf 07} (2007) 062,
  [\href{http://xxx.lanl.gov/abs/0705.0016}{{\tt arXiv:0705.0016}}].

\bibitem{PhysRevLett.28.1082}
J.~W. York, {\it Role of conformal three-geometry in the dynamics of
  gravitation},  {\em Phys. Rev. Lett.} {\bf 28} (Apr, 1972) 1082--1085.

\bibitem{PhysRevD.15.2752}
G.~W. Gibbons and S.~W. Hawking, {\it Action integrals and partition functions
  in quantum gravity},  {\em Phys. Rev. D} {\bf 15} (May, 1977) 2752--2756.

\bibitem{Dabholkar:1994ai}
A.~Dabholkar, {\it {Strings on a cone and black hole entropy}},  {\em Nucl.
  Phys. B} {\bf 439} (1995) 650--664,
  [\href{http://xxx.lanl.gov/abs/hep-th/9408098}{{\tt hep-th/9408098}}].

\bibitem{Dabholkar:2001if}
A.~Dabholkar, {\it {Tachyon condensation and black hole entropy}},  {\em Phys.
  Rev. Lett.} {\bf 88} (2002) 091301,
  [\href{http://xxx.lanl.gov/abs/hep-th/0111004}{{\tt hep-th/0111004}}].

\bibitem{Sen:2008vm}
A.~Sen, {\it {Quantum Entropy Function from AdS(2)/CFT(1) Correspondence}},
  {\em Int. J. Mod. Phys. A} {\bf 24} (2009) 4225--4244,
  [\href{http://xxx.lanl.gov/abs/0809.3304}{{\tt arXiv:0809.3304}}].

\bibitem{Lewkowycz:2013nqa}
A.~Lewkowycz and J.~Maldacena, {\it {Generalized gravitational entropy}},  {\em
  JHEP} {\bf 08} (2013) 090, [\href{http://xxx.lanl.gov/abs/1304.4926}{{\tt
  arXiv:1304.4926}}].

\bibitem{Faulkner:2013ana}
T.~Faulkner, A.~Lewkowycz, and J.~Maldacena, {\it {Quantum corrections to
  holographic entanglement entropy}},  {\em JHEP} {\bf 11} (2013) 074,
  [\href{http://xxx.lanl.gov/abs/1307.2892}{{\tt arXiv:1307.2892}}].

\bibitem{Engelhardt:2014gca}
N.~Engelhardt and A.~C. Wall, {\it {Quantum Extremal Surfaces: Holographic
  Entanglement Entropy beyond the Classical Regime}},  {\em JHEP} {\bf 01}
  (2015) 073, [\href{http://xxx.lanl.gov/abs/1408.3203}{{\tt
  arXiv:1408.3203}}].

\bibitem{Maldacena:1998bw}
J.~M. Maldacena and A.~Strominger, {\it {AdS(3) black holes and a stringy
  exclusion principle}},  {\em JHEP} {\bf 12} (1998) 005,
  [\href{http://xxx.lanl.gov/abs/hep-th/9804085}{{\tt hep-th/9804085}}].

\bibitem{Eberhardt:2023lwd}
L.~Eberhardt and S.~Pal, {\it {Holographic Weyl anomaly in string theory}},
  [\href{http://xxx.lanl.gov/abs/2307.03000}{{\tt arXiv:2307.03000}}].

\bibitem{Liu:1987nz}
J.~Liu and J.~Polchinski, {\it {Renormalization of the Mobius Volume}},  {\em
  Phys. Lett. B} {\bf 203} (1988) 39--43.

\bibitem{Polchinski:1995mt}
J.~Polchinski, {\it {Dirichlet Branes and Ramond-Ramond charges}},  {\em Phys.
  Rev. Lett.} {\bf 75} (1995) 4724--4727,
  [\href{http://xxx.lanl.gov/abs/hep-th/9510017}{{\tt hep-th/9510017}}].

\bibitem{Eberhardt:2021ynh}
L.~Eberhardt and S.~Pal, {\it {The disk partition function in string theory}},
  {\em JHEP} {\bf 08} (2021) 026,
  [\href{http://xxx.lanl.gov/abs/2105.08726}{{\tt arXiv:2105.08726}}].

\bibitem{Erbin:2019uiz}
H.~Erbin, J.~Maldacena, and D.~Skliros, {\it {Two-Point String Amplitudes}},
  {\em JHEP} {\bf 07} (2019) 139,
  [\href{http://xxx.lanl.gov/abs/1906.06051}{{\tt arXiv:1906.06051}}].

\bibitem{Troost}
J.~Troost, {\it {The $AdS_3$ central charge in string theory}},  {\em Phys.
  Lett. B} {\bf 705} (2011) 260--263,
  [\href{http://xxx.lanl.gov/abs/1109.1923}{{\tt arXiv:1109.1923}}].

\bibitem{Jafferis:2021ywg}
D.~L. Jafferis and E.~Schneider, {\it {Stringy ER=EPR}},
  [\href{http://xxx.lanl.gov/abs/2104.07233}{{\tt arXiv:2104.07233}}].

\bibitem{Halder:2022ykw}
I.~Halder, D.~L. Jafferis, and D.~Kolchmeyer, {\it {A duality in string theory
  on AdS$_3$}},  [\href{http://xxx.lanl.gov/abs/2208.00016}{{\tt
  arXiv:2208.00016}}].

\bibitem{Halder:2023nlp}
I.~Halder and D.~L. Jafferis, {\it {Double winding condensate CFT}},
  [\href{http://xxx.lanl.gov/abs/2308.11702}{{\tt arXiv:2308.11702}}].

\bibitem{Erler:2022agw}
T.~Erler, {\it {The closed string field theory action vanishes}},  {\em JHEP}
  {\bf 10} (2022) 055, [\href{http://xxx.lanl.gov/abs/2204.12863}{{\tt
  arXiv:2204.12863}}].

\bibitem{Bergman:1994qq}
O.~Bergman and B.~Zwiebach, {\it {The Dilaton theorem and closed string
  backgrounds}},  {\em Nucl. Phys. B} {\bf 441} (1995) 76--118,
  [\href{http://xxx.lanl.gov/abs/hep-th/9411047}{{\tt hep-th/9411047}}].

\bibitem{Adams:2001sv}
A.~Adams, J.~Polchinski, and E.~Silverstein, {\it {Don't panic! Closed string
  tachyons in ALE space-times}},  {\em JHEP} {\bf 10} (2001) 029,
  [\href{http://xxx.lanl.gov/abs/hep-th/0108075}{{\tt hep-th/0108075}}].

\bibitem{Okawa:2004rh}
Y.~Okawa and B.~Zwiebach, {\it {Twisted tachyon condensation in closed string
  field theory}},  {\em JHEP} {\bf 03} (2004) 056,
  [\href{http://xxx.lanl.gov/abs/hep-th/0403051}{{\tt hep-th/0403051}}].

\bibitem{Headrick:2004hz}
M.~Headrick, S.~Minwalla, and T.~Takayanagi, {\it {Closed string tachyon
  condensation: An Overview}},  {\em Class. Quant. Grav.} {\bf 21} (2004)
  S1539--S1565, [\href{http://xxx.lanl.gov/abs/hep-th/0405064}{{\tt
  hep-th/0405064}}].

\bibitem{Tseytlin:1988tv}
A.~A. Tseytlin, {\it {Mobius Infinity Subtraction and Effective Action in
  $\sigma$ Model Approach to Closed String Theory}},  {\em Phys. Lett. B} {\bf
  208} (1988) 221--227.

\bibitem{Tseytlin:1988rr}
A.~A. Tseytlin, {\it {SIGMA MODEL APPROACH TO STRING THEORY}},  {\em Int. J.
  Mod. Phys. A} {\bf 4} (1989) 1257.

\bibitem{Osborn:1989bu}
H.~Osborn, {\it {General Bosonic $\sigma$ Models and String Effective
  Actions}},  {\em Annals Phys.} {\bf 200} (1990) 1.

\bibitem{Osborn:1991gm}
H.~Osborn, {\it {Weyl consistency conditions and a local renormalization group
  equation for general renormalizable field theories}},  {\em Nucl. Phys. B}
  {\bf 363} (1991) 486--526.

\bibitem{Carlip:1993sa}
S.~Carlip and C.~Teitelboim, {\it {The Off-shell black hole}},  {\em Class.
  Quant. Grav.} {\bf 12} (1995) 1699--1704,
  [\href{http://xxx.lanl.gov/abs/gr-qc/9312002}{{\tt gr-qc/9312002}}].

\bibitem{Banados:1993qp}
M.~Banados, C.~Teitelboim, and J.~Zanelli, {\it {Black hole entropy and the
  dimensional continuation of the Gauss-Bonnet theorem}},  {\em Phys. Rev.
  Lett.} {\bf 72} (1994) 957--960,
  [\href{http://xxx.lanl.gov/abs/gr-qc/9309026}{{\tt gr-qc/9309026}}].

\bibitem{Susskind:1994sm}
L.~Susskind and J.~Uglum, {\it {Black hole entropy in canonical quantum gravity
  and superstring theory}},  {\em Phys. Rev. D} {\bf 50} (1994) 2700--2711,
  [\href{http://xxx.lanl.gov/abs/hep-th/9401070}{{\tt hep-th/9401070}}].

\bibitem{Fursaev:2006ih}
D.~V. Fursaev, {\it {Proof of the holographic formula for entanglement
  entropy}},  {\em JHEP} {\bf 09} (2006) 018,
  [\href{http://xxx.lanl.gov/abs/hep-th/0606184}{{\tt hep-th/0606184}}].

\bibitem{Headrick:2010zt}
M.~Headrick, {\it {Entanglement Renyi entropies in holographic theories}},
  {\em Phys. Rev. D} {\bf 82} (2010) 126010,
  [\href{http://xxx.lanl.gov/abs/1006.0047}{{\tt arXiv:1006.0047}}].

\bibitem{Brustein:2021cza}
R.~Brustein and Y.~Zigdon, {\it {Black hole entropy sourced by string winding
  condensate}},  {\em JHEP} {\bf 10} (2021) 219,
  [\href{http://xxx.lanl.gov/abs/2107.09001}{{\tt arXiv:2107.09001}}].

\bibitem{Ahmadain:2022tew}
A.~Ahmadain and A.~C. Wall, {\it {Off-Shell Strings I: S-matrix and Action}},
  [\href{http://xxx.lanl.gov/abs/2211.08607}{{\tt arXiv:2211.08607}}].

\bibitem{Ahmadain:2022eso}
A.~Ahmadain and A.~C. Wall, {\it {Off-Shell Strings II: Black Hole Entropy}},
  [\href{http://xxx.lanl.gov/abs/2211.16448}{{\tt arXiv:2211.16448}}].

\bibitem{Witten:2018xfj}
E.~Witten, {\it {Open Strings On The Rindler Horizon}},  {\em JHEP} {\bf 01}
  (2019) 126, [\href{http://xxx.lanl.gov/abs/1810.11912}{{\tt
  arXiv:1810.11912}}].

\bibitem{Dabholkar:2022mxo}
A.~Dabholkar, {\it {Quantum Entanglement in String Theory}},
  [\href{http://xxx.lanl.gov/abs/2207.03624}{{\tt arXiv:2207.03624}}].

\bibitem{He:2014gva}
S.~He, T.~Numasawa, T.~Takayanagi, and K.~Watanabe, {\it {Notes on Entanglement
  Entropy in String Theory}},  {\em JHEP} {\bf 05} (2015) 106,
  [\href{http://xxx.lanl.gov/abs/1412.5606}{{\tt arXiv:1412.5606}}].

\bibitem{Ribault:2005wp}
S.~Ribault and J.~Teschner, {\it {H+(3)-WZNW correlators from Liouville
  theory}},  {\em JHEP} {\bf 06} (2005) 014,
  [\href{http://xxx.lanl.gov/abs/hep-th/0502048}{{\tt hep-th/0502048}}].

\bibitem{Giveon:1998ns}
A.~Giveon, D.~Kutasov, and N.~Seiberg, {\it {Comments on string theory on
  AdS(3)}},  {\em Adv. Theor. Math. Phys.} {\bf 2} (1998) 733--782,
  [\href{http://xxx.lanl.gov/abs/hep-th/9806194}{{\tt hep-th/9806194}}].

\bibitem{Kutasov:1999xu}
D.~Kutasov and N.~Seiberg, {\it {More comments on string theory on AdS(3)}},
  {\em JHEP} {\bf 04} (1999) 008,
  [\href{http://xxx.lanl.gov/abs/hep-th/9903219}{{\tt hep-th/9903219}}].

\bibitem{Teschner:1997ft}
J.~Teschner, {\it {On structure constants and fusion rules in the SL(2,C) /
  SU(2) WZNW model}},  {\em Nucl. Phys. B} {\bf 546} (1999) 390--422,
  [\href{http://xxx.lanl.gov/abs/hep-th/9712256}{{\tt hep-th/9712256}}].

\bibitem{Teschner:1999ug}
J.~Teschner, {\it {Operator product expansion and factorization in the H+(3)
  WZNW model}},  {\em Nucl. Phys. B} {\bf 571} (2000) 555--582,
  [\href{http://xxx.lanl.gov/abs/hep-th/9906215}{{\tt hep-th/9906215}}].

\bibitem{Maldacena:2000hw}
J.~M. Maldacena and H.~Ooguri, {\it {Strings in AdS(3) and SL(2,R) WZW model
  1.: The Spectrum}},  {\em J. Math. Phys.} {\bf 42} (2001) 2929--2960,
  [\href{http://xxx.lanl.gov/abs/hep-th/0001053}{{\tt hep-th/0001053}}].

\bibitem{Maldacena:2000kv}
J.~M. Maldacena, H.~Ooguri, and J.~Son, {\it {Strings in AdS(3) and the SL(2,R)
  WZW model. Part 2. Euclidean black hole}},  {\em J. Math. Phys.} {\bf 42}
  (2001) 2961--2977, [\href{http://xxx.lanl.gov/abs/hep-th/0005183}{{\tt
  hep-th/0005183}}].

\bibitem{Maldacena:2001km}
J.~M. Maldacena and H.~Ooguri, {\it {Strings in AdS(3) and the SL(2,R) WZW
  model. Part 3. Correlation functions}},  {\em Phys. Rev. D} {\bf 65} (2002)
  106006, [\href{http://xxx.lanl.gov/abs/hep-th/0111180}{{\tt
  hep-th/0111180}}].

\bibitem{Dei:2021yom}
A.~Dei and L.~Eberhardt, {\it {String correlators on AdS$_{3}$: four-point
  functions}},  {\em JHEP} {\bf 09} (2021) 209,
  [\href{http://xxx.lanl.gov/abs/2107.01481}{{\tt arXiv:2107.01481}}].

\bibitem{Dei:2021xgh}
A.~Dei and L.~Eberhardt, {\it {String correlators on AdS$_{3}$: three-point
  functions}},  {\em JHEP} {\bf 08} (2021) 025,
  [\href{http://xxx.lanl.gov/abs/2105.12130}{{\tt arXiv:2105.12130}}].

\bibitem{Gaberdiel:2018rqv}
M.~R. Gaberdiel and R.~Gopakumar, {\it {Tensionless string spectra on
  AdS$_{3}$}},  {\em JHEP} {\bf 05} (2018) 085,
  [\href{http://xxx.lanl.gov/abs/1803.04423}{{\tt arXiv:1803.04423}}].

\bibitem{Eberhardt:2018ouy}
L.~Eberhardt, M.~R. Gaberdiel, and R.~Gopakumar, {\it {The Worldsheet Dual of
  the Symmetric Product CFT}},  {\em JHEP} {\bf 04} (2019) 103,
  [\href{http://xxx.lanl.gov/abs/1812.01007}{{\tt arXiv:1812.01007}}].

\bibitem{Eberhardt:2019ywk}
L.~Eberhardt, M.~R. Gaberdiel, and R.~Gopakumar, {\it {Deriving the
  AdS$_{3}$/CFT$_{2}$ correspondence}},  {\em JHEP} {\bf 02} (2020) 136,
  [\href{http://xxx.lanl.gov/abs/1911.00378}{{\tt arXiv:1911.00378}}].

\bibitem{Dei:2020zui}
A.~Dei, M.~R. Gaberdiel, R.~Gopakumar, and B.~Knighton, {\it {Free field
  world-sheet correlators for ${\rm AdS}_3$}},  {\em JHEP} {\bf 02} (2021) 081,
  [\href{http://xxx.lanl.gov/abs/2009.11306}{{\tt arXiv:2009.11306}}].

\bibitem{Zamolodchikov:2001dz}
A.~Zamolodchikov, {\it {Scaling Lee-Yang model on a sphere. 1. Partition
  function}},  {\em JHEP} {\bf 07} (2002) 029,
  [\href{http://xxx.lanl.gov/abs/hep-th/0109078}{{\tt hep-th/0109078}}].

\bibitem{Zamolodchikov:1995aa}
A.~B. Zamolodchikov and A.~B. Zamolodchikov, {\it {Structure constants and
  conformal bootstrap in Liouville field theory}},  {\em Nucl. Phys. B} {\bf
  477} (1996) 577--605, [\href{http://xxx.lanl.gov/abs/hep-th/9506136}{{\tt
  hep-th/9506136}}].

\bibitem{Polchinski:v1}
J.~Polchinski, {\em {String theory. Vol. 1: An introduction to the bosonic
  string}}.
\newblock Cambridge Monographs on Mathematical Physics. Cambridge University
  Press, 12, 2007.

\bibitem{Frenkel:2005ku}
E.~Frenkel and A.~Losev, {\it {Mirror symmetry in two steps: A-I-B}},  {\em
  Commun. Math. Phys.} {\bf 269} (2006) 39--86,
  [\href{http://xxx.lanl.gov/abs/hep-th/0505131}{{\tt hep-th/0505131}}].

\bibitem{PhysRevLett.66.2051}
M.~Goulian and M.~Li, {\it Correlation functions in liouville theory},  {\em
  Phys. Rev. Lett.} {\bf 66} (Apr, 1991) 2051--2055.

\bibitem{Dorn:1994xn}
H.~Dorn and H.~J. Otto, {\it {Two and three point functions in Liouville
  theory}},  {\em Nucl. Phys. B} {\bf 429} (1994) 375--388,
  [\href{http://xxx.lanl.gov/abs/hep-th/9403141}{{\tt hep-th/9403141}}].

\bibitem{Teschner:2001rv}
J.~Teschner, {\it {Liouville theory revisited}},  {\em Class. Quant. Grav.}
  {\bf 18} (2001) R153--R222,
  [\href{http://xxx.lanl.gov/abs/hep-th/0104158}{{\tt hep-th/0104158}}].

\bibitem{Balthazar:2021xeh}
B.~Balthazar, A.~Giveon, D.~Kutasov, and E.~J. Martinec, {\it {Asymptotically
  free AdS$_{3}$/CFT$_{2}$}},  {\em JHEP} {\bf 01} (2022) 008,
  [\href{http://xxx.lanl.gov/abs/2109.00065}{{\tt arXiv:2109.00065}}].

\bibitem{Mahajan:2021nsd}
R.~Mahajan, D.~Stanford, and C.~Yan, {\it {Sphere and disk partition functions
  in Liouville and in matrix integrals}},  {\em JHEP} {\bf 07} (2022) 132,
  [\href{http://xxx.lanl.gov/abs/2107.01172}{{\tt arXiv:2107.01172}}].

\bibitem{Giribet:2007uh}
G.~Giribet and M.~Leoni, {\it {A Twisted FZZ-like dual for the 2D black hole}},
   {\em Rept. Math. Phys.} {\bf 61} (2008) 151--162,
  [\href{http://xxx.lanl.gov/abs/0706.0036}{{\tt arXiv:0706.0036}}].

\bibitem{Fateev:2007qn}
V.~A. Fateev and A.~V. Litvinov, {\it {Multipoint correlation functions in
  Liouville field theory and minimal Liouville gravity}},  {\em Theor. Math.
  Phys.} {\bf 154} (2008) 454--472,
  [\href{http://xxx.lanl.gov/abs/0707.1664}{{\tt arXiv:0707.1664}}].

\bibitem{Eberhardt:2021vsx}
L.~Eberhardt, {\it {A perturbative CFT dual for pure NS\textendash{}NS
  AdS$_{3}$ strings}},  {\em J. Phys. A} {\bf 55} (2022), no.~6 064001,
  [\href{http://xxx.lanl.gov/abs/2110.07535}{{\tt arXiv:2110.07535}}].

\bibitem{Polchinski:1985zf}
J.~Polchinski, {\it {Evaluation of the One Loop String Path Integral}},  {\em
  Commun. Math. Phys.} {\bf 104} (1986) 37.

\bibitem{Halder:2024gwe}
I.~Halder and D.~L. Jafferis, {\it {Stretched horizon, replica trick and
  off-shell winding condensate, and all that}},
  [\href{http://xxx.lanl.gov/abs/2402.00932}{{\tt arXiv:2402.00932}}].

\bibitem{Kraus:2002cb}
P.~Kraus, A.~Ryzhov, and M.~Shigemori, {\it {Strings in noncompact space-times:
  Boundary terms and conserved charges}},  {\em Phys. Rev. D} {\bf 66} (2002)
  106001, [\href{http://xxx.lanl.gov/abs/hep-th/0206080}{{\tt
  hep-th/0206080}}].

\bibitem{Kraus:2005vz}
P.~Kraus and F.~Larsen, {\it {Microscopic black hole entropy in theories with
  higher derivatives}},  {\em JHEP} {\bf 09} (2005) 034,
  [\href{http://xxx.lanl.gov/abs/hep-th/0506176}{{\tt hep-th/0506176}}].

\bibitem{Kraus:2005zm}
P.~Kraus and F.~Larsen, {\it {Holographic gravitational anomalies}},  {\em
  JHEP} {\bf 01} (2006) 022,
  [\href{http://xxx.lanl.gov/abs/hep-th/0508218}{{\tt hep-th/0508218}}].

\bibitem{David:2007ak}
J.~R. David, B.~Sahoo, and A.~Sen, {\it {AdS(3), black holes and higher
  derivative corrections}},  {\em JHEP} {\bf 07} (2007) 058,
  [\href{http://xxx.lanl.gov/abs/0705.0735}{{\tt arXiv:0705.0735}}].

\bibitem{Creutzig:2013hma}
T.~Creutzig and D.~Ridout, {\it {Logarithmic Conformal Field Theory: Beyond an
  Introduction}},  {\em J. Phys. A} {\bf 46} (2013) 4006,
  [\href{http://xxx.lanl.gov/abs/1303.0847}{{\tt arXiv:1303.0847}}].

\bibitem{Leutheusser:2021qhd}
S.~Leutheusser and H.~Liu, {\it {Causal connectability between quantum systems
  and the black hole interior in holographic duality}},
  [\href{http://xxx.lanl.gov/abs/2110.05497}{{\tt arXiv:2110.05497}}].

\bibitem{Leutheusser:2021frk}
S.~Leutheusser and H.~Liu, {\it {Emergent times in holographic duality}},
  [\href{http://xxx.lanl.gov/abs/2112.12156}{{\tt arXiv:2112.12156}}].

\bibitem{Witten:2021unn}
E.~Witten, {\it {Gravity and the crossed product}},  {\em JHEP} {\bf 10} (2022)
  008, [\href{http://xxx.lanl.gov/abs/2112.12828}{{\tt arXiv:2112.12828}}].

\bibitem{Chandrasekaran:2022eqq}
V.~Chandrasekaran, G.~Penington, and E.~Witten, {\it {Large N algebras and
  generalized entropy}},  [\href{http://xxx.lanl.gov/abs/2209.10454}{{\tt
  arXiv:2209.10454}}].

\bibitem{Verlinde:2020upt}
H.~Verlinde, {\it {ER = EPR revisited: On the Entropy of an Einstein-Rosen
  Bridge}},  [\href{http://xxx.lanl.gov/abs/2003.13117}{{\tt
  arXiv:2003.13117}}].

\bibitem{Strominger:1996sh}
A.~Strominger and C.~Vafa, {\it {Microscopic origin of the Bekenstein-Hawking
  entropy}},  {\em Phys. Lett. B} {\bf 379} (1996) 99--104,
  [\href{http://xxx.lanl.gov/abs/hep-th/9601029}{{\tt hep-th/9601029}}].

\bibitem{Sen:1995in}
A.~Sen, {\it {Extremal black holes and elementary string states}},  {\em Mod.
  Phys. Lett. A} {\bf 10} (1995) 2081--2094,
  [\href{http://xxx.lanl.gov/abs/hep-th/9504147}{{\tt hep-th/9504147}}].

\bibitem{Sen:2004dp}
A.~Sen, {\it {How does a fundamental string stretch its horizon?}},  {\em JHEP}
  {\bf 05} (2005) 059, [\href{http://xxx.lanl.gov/abs/hep-th/0411255}{{\tt
  hep-th/0411255}}].

\bibitem{Gopakumar:1998ii}
R.~Gopakumar and C.~Vafa, {\it {M theory and topological strings. 1.}},
  [\href{http://xxx.lanl.gov/abs/hep-th/9809187}{{\tt hep-th/9809187}}].

\bibitem{Gopakumar:1998jq}
R.~Gopakumar and C.~Vafa, {\it {M theory and topological strings. 2.}},
  [\href{http://xxx.lanl.gov/abs/hep-th/9812127}{{\tt hep-th/9812127}}].

\bibitem{HalderLin}
I.~Halder and Y.-H. Lin, {\it {Blackhole/blackring transition}},
  [\href{http://xxx.lanl.gov/abs/2307.13735}{{\tt arXiv:2307.13735}}].

\bibitem{Kinney:2005ej}
J.~Kinney, J.~M. Maldacena, S.~Minwalla, and S.~Raju, {\it {An Index for 4
  dimensional super conformal theories}},  {\em Commun. Math. Phys.} {\bf 275}
  (2007) 209--254, [\href{http://xxx.lanl.gov/abs/hep-th/0510251}{{\tt
  hep-th/0510251}}].

\bibitem{Bhattacharya:2008zy}
J.~Bhattacharya, S.~Bhattacharyya, S.~Minwalla, and S.~Raju, {\it {Indices for
  Superconformal Field Theories in 3,5 and 6 Dimensions}},  {\em JHEP} {\bf 02}
  (2008) 064, [\href{http://xxx.lanl.gov/abs/0801.1435}{{\tt
  arXiv:0801.1435}}].

\bibitem{Ooguri:2004zv}
H.~Ooguri, A.~Strominger, and C.~Vafa, {\it {Black hole attractors and the
  topological string}},  {\em Phys. Rev. D} {\bf 70} (2004) 106007,
  [\href{http://xxx.lanl.gov/abs/hep-th/0405146}{{\tt hep-th/0405146}}].

\bibitem{Denef:2007vg}
F.~Denef and G.~W. Moore, {\it {Split states, entropy enigmas, holes and
  halos}},  {\em JHEP} {\bf 11} (2011) 129,
  [\href{http://xxx.lanl.gov/abs/hep-th/0702146}{{\tt hep-th/0702146}}].

\bibitem{Benini:2015eyy}
F.~Benini, K.~Hristov, and A.~Zaffaroni, {\it {Black hole microstates in
  AdS$_{4}$ from supersymmetric localization}},  {\em JHEP} {\bf 05} (2016)
  054, [\href{http://xxx.lanl.gov/abs/1511.04085}{{\tt arXiv:1511.04085}}].

\bibitem{Cabo-Bizet:2018ehj}
A.~Cabo-Bizet, D.~Cassani, D.~Martelli, and S.~Murthy, {\it {Microscopic origin
  of the Bekenstein-Hawking entropy of supersymmetric AdS$_{5}$ black holes}},
  {\em JHEP} {\bf 10} (2019) 062,
  [\href{http://xxx.lanl.gov/abs/1810.11442}{{\tt arXiv:1810.11442}}].

\bibitem{Choi:2018hmj}
S.~Choi, J.~Kim, S.~Kim, and J.~Nahmgoong, {\it {Large AdS black holes from
  QFT}},  [\href{http://xxx.lanl.gov/abs/1810.12067}{{\tt arXiv:1810.12067}}].

\bibitem{Iliesiu:2021are}
L.~V. Iliesiu, M.~Kologlu, and G.~J. Turiaci, {\it {Supersymmetric indices
  factorize}},  {\em JHEP} {\bf 05} (2023) 032,
  [\href{http://xxx.lanl.gov/abs/2107.09062}{{\tt arXiv:2107.09062}}].

\bibitem{H:2023qko}
A.~A. H., P.~V. Athira, C.~Chowdhury, and A.~Sen, {\it {Logarithmic Correction
  to BPS Black Hole Entropy from Supersymmetric Index at Finite Temperature}},
  [\href{http://xxx.lanl.gov/abs/2306.07322}{{\tt arXiv:2306.07322}}].

\bibitem{Chen:2023mbc}
Y.~Chen and G.~J. Turiaci, {\it {Spin-Statistics for Black Hole Microstates}},
  [\href{http://xxx.lanl.gov/abs/2309.03478}{{\tt arXiv:2309.03478}}].

\bibitem{Chen:2023lzq}
Y.~Chen, M.~Heydeman, Y.~Wang, and M.~Zhang, {\it {Probing Supersymmetric Black
  Holes with Surface Defects}},
  [\href{http://xxx.lanl.gov/abs/2306.05463}{{\tt arXiv:2306.05463}}].

\bibitem{Horowitz:1996ay}
G.~T. Horowitz, J.~M. Maldacena, and A.~Strominger, {\it {Nonextremal black
  hole microstates and U duality}},  {\em Phys. Lett. B} {\bf 383} (1996)
  151--159, [\href{http://xxx.lanl.gov/abs/hep-th/9603109}{{\tt
  hep-th/9603109}}].

\bibitem{Gubser:1996de}
S.~S. Gubser, I.~R. Klebanov, and A.~W. Peet, {\it {Entropy and temperature of
  black 3-branes}},  {\em Phys. Rev. D} {\bf 54} (1996) 3915--3919,
  [\href{http://xxx.lanl.gov/abs/hep-th/9602135}{{\tt hep-th/9602135}}].

\bibitem{Biggs:2023sqw}
A.~Biggs and J.~Maldacena, {\it {Scaling similarities and quasinormal modes of
  D0 black hole solutions}},  [\href{http://xxx.lanl.gov/abs/2303.09974}{{\tt
  arXiv:2303.09974}}].

\bibitem{Paban:1998ea}
S.~Paban, S.~Sethi, and M.~Stern, {\it {Constraints from extended supersymmetry
  in quantum mechanics}},  {\em Nucl. Phys. B} {\bf 534} (1998) 137--154,
  [\href{http://xxx.lanl.gov/abs/hep-th/9805018}{{\tt hep-th/9805018}}].

\bibitem{Paban:1998qy}
S.~Paban, S.~Sethi, and M.~Stern, {\it {Supersymmetry and higher derivative
  terms in the effective action of Yang-Mills theories}},  {\em JHEP} {\bf 06}
  (1998) 012, [\href{http://xxx.lanl.gov/abs/hep-th/9806028}{{\tt
  hep-th/9806028}}].

\bibitem{Heydeman:2020hhw}
M.~Heydeman, L.~V. Iliesiu, G.~J. Turiaci, and W.~Zhao, {\it {The statistical
  mechanics of near-BPS black holes}},  {\em J. Phys. A} {\bf 55} (2022), no.~1
  014004, [\href{http://xxx.lanl.gov/abs/2011.01953}{{\tt arXiv:2011.01953}}].

\bibitem{Iliesiu:2020qvm}
L.~V. Iliesiu and G.~J. Turiaci, {\it {The statistical mechanics of
  near-extremal black holes}},  {\em JHEP} {\bf 05} (2021) 145,
  [\href{http://xxx.lanl.gov/abs/2003.02860}{{\tt arXiv:2003.02860}}].

\bibitem{Maldacena:2016hyu}
J.~Maldacena and D.~Stanford, {\it {Remarks on the Sachdev-Ye-Kitaev model}},
  {\em Phys. Rev. D} {\bf 94} (2016), no.~10 106002,
  [\href{http://xxx.lanl.gov/abs/1604.07818}{{\tt arXiv:1604.07818}}].

\bibitem{Maldacena:2015waa}
J.~Maldacena, S.~H. Shenker, and D.~Stanford, {\it {A bound on chaos}},  {\em
  JHEP} {\bf 08} (2016) 106, [\href{http://xxx.lanl.gov/abs/1503.01409}{{\tt
  arXiv:1503.01409}}].

\bibitem{Halder:2019ric}
I.~Halder, {\it {Global symmetry and maximal chaos}},
  [\href{http://xxx.lanl.gov/abs/1908.05281}{{\tt arXiv:1908.05281}}].

\bibitem{Camanho:2014apa}
X.~O. Camanho, J.~D. Edelstein, J.~Maldacena, and A.~Zhiboedov, {\it {Causality
  constraints on corrections to the graviton three-point coupling}},  {\em
  JHEP} {\bf 02} (2016) 020, [\href{http://xxx.lanl.gov/abs/1407.5597}{{\tt
  arXiv:1407.5597}}].

\bibitem{Chowdhury:2019kaq}
S.~D. Chowdhury, A.~Gadde, T.~Gopalka, I.~Halder, L.~Janagal, and S.~Minwalla,
  {\it {Classifying and constraining local four photon and four graviton
  S-matrices}},  {\em JHEP} {\bf 02} (2020) 114,
  [\href{http://xxx.lanl.gov/abs/1910.14392}{{\tt arXiv:1910.14392}}].

\bibitem{Chandorkar:2021viw}
D.~Chandorkar, S.~D. Chowdhury, S.~Kundu, and S.~Minwalla, {\it {Bounds on
  Regge growth of flat space scattering from bounds on chaos}},  {\em JHEP}
  {\bf 05} (2021) 143, [\href{http://xxx.lanl.gov/abs/2102.03122}{{\tt
  arXiv:2102.03122}}].

\bibitem{Blake:2021wqj}
M.~Blake and H.~Liu, {\it {On systems of maximal quantum chaos}},  {\em JHEP}
  {\bf 05} (2021) 229, [\href{http://xxx.lanl.gov/abs/2102.11294}{{\tt
  arXiv:2102.11294}}].

\bibitem{Das:1996wn}
S.~R. Das and S.~D. Mathur, {\it {Comparing decay rates for black holes and
  D-branes}},  {\em Nucl. Phys. B} {\bf 478} (1996) 561--576,
  [\href{http://xxx.lanl.gov/abs/hep-th/9606185}{{\tt hep-th/9606185}}].

\bibitem{Das:1996jy}
S.~R. Das and S.~D. Mathur, {\it {Interactions involving D-branes}},  {\em
  Nucl. Phys. B} {\bf 482} (1996) 153--172,
  [\href{http://xxx.lanl.gov/abs/hep-th/9607149}{{\tt hep-th/9607149}}].

\bibitem{Dhar:1996vu}
A.~Dhar, G.~Mandal, and S.~R. Wadia, {\it {Absorption versus decay of black
  holes in string theory and T symmetry}},  {\em Phys. Lett. B} {\bf 388}
  (1996) 51--59, [\href{http://xxx.lanl.gov/abs/hep-th/9605234}{{\tt
  hep-th/9605234}}].

\bibitem{Maldacena:1997ih}
J.~M. Maldacena and A.~Strominger, {\it {Universal low-energy dynamics for
  rotating black holes}},  {\em Phys. Rev. D} {\bf 56} (1997) 4975--4983,
  [\href{http://xxx.lanl.gov/abs/hep-th/9702015}{{\tt hep-th/9702015}}].

\bibitem{Agia:2022srj}
N.~Agia and D.~L. Jafferis, {\it {Angular Quantization in CFT}},
  [\href{http://xxx.lanl.gov/abs/2204.11872}{{\tt arXiv:2204.11872}}].

\bibitem{Dijkgraaf:1991ba}
R.~Dijkgraaf, H.~L. Verlinde, and E.~P. Verlinde, {\it {String propagation in a
  black hole geometry}},  {\em Nucl. Phys. B} {\bf 371} (1992) 269--314.

\bibitem{Mertens:2013zya}
T.~G. Mertens, H.~Verschelde, and V.~I. Zakharov, {\it {Random Walks in Rindler
  Spacetime and String Theory at the Tip of the Cigar}},  {\em JHEP} {\bf 03}
  (2014) 086, [\href{http://xxx.lanl.gov/abs/1307.3491}{{\tt
  arXiv:1307.3491}}].

\bibitem{Almheiri:2020cfm}
A.~Almheiri, T.~Hartman, J.~Maldacena, E.~Shaghoulian, and A.~Tajdini, {\it
  {The entropy of Hawking radiation}},  {\em Rev. Mod. Phys.} {\bf 93} (2021),
  no.~3 035002, [\href{http://xxx.lanl.gov/abs/2006.06872}{{\tt
  arXiv:2006.06872}}].

\bibitem{Raju:2020smc}
S.~Raju, {\it {Lessons from the information paradox}},  {\em Phys. Rept.} {\bf
  943} (2022) 1--80, [\href{http://xxx.lanl.gov/abs/2012.05770}{{\tt
  arXiv:2012.05770}}].

\bibitem{Hartnoll:2015fca}
S.~A. Hartnoll and E.~Mazenc, {\it {Entanglement entropy in two dimensional
  string theory}},  {\em Phys. Rev. Lett.} {\bf 115} (2015), no.~12 121602,
  [\href{http://xxx.lanl.gov/abs/1504.07985}{{\tt arXiv:1504.07985}}].

\bibitem{Das:2020jhy}
S.~R. Das, A.~Kaushal, G.~Mandal, and S.~P. Trivedi, {\it {Bulk Entanglement
  Entropy and Matrices}},  {\em J. Phys. A} {\bf 53} (2020), no.~44 444002,
  [\href{http://xxx.lanl.gov/abs/2004.00613}{{\tt arXiv:2004.00613}}].

\bibitem{Das:2020xoa}
S.~R. Das, A.~Kaushal, S.~Liu, G.~Mandal, and S.~P. Trivedi, {\it {Gauge
  invariant target space entanglement in D-brane holography}},  {\em JHEP} {\bf
  04} (2021) 225, [\href{http://xxx.lanl.gov/abs/2011.13857}{{\tt
  arXiv:2011.13857}}].

\bibitem{Das:2022njy}
S.~R. Das, A.~Kaushal, G.~Mandal, K.~K. Nanda, M.~H. Radwan, and S.~P. Trivedi,
  {\it {Entanglement entropy in internal spaces and Ryu-Takayanagi surfaces}},
  {\em JHEP} {\bf 04} (2023) 141,
  [\href{http://xxx.lanl.gov/abs/2212.11640}{{\tt arXiv:2212.11640}}].

\bibitem{Balasubramanian:2018axm}
V.~Balasubramanian and O.~Parrikar, {\it {Remarks on entanglement entropy in
  string theory}},  {\em Phys. Rev. D} {\bf 97} (2018), no.~6 066025,
  [\href{http://xxx.lanl.gov/abs/1801.03517}{{\tt arXiv:1801.03517}}].

\bibitem{Giveon:2016dxe}
A.~Giveon, N.~Itzhaki, and D.~Kutasov, {\it {Stringy Horizons II}},  {\em JHEP}
  {\bf 10} (2016) 157, [\href{http://xxx.lanl.gov/abs/1603.05822}{{\tt
  arXiv:1603.05822}}].

\bibitem{Martinec:2020gkv}
E.~J. Martinec, S.~Massai, and D.~Turton, {\it {Stringy Structure at the BPS
  Bound}},  {\em JHEP} {\bf 12} (2020) 135,
  [\href{http://xxx.lanl.gov/abs/2005.12344}{{\tt arXiv:2005.12344}}].

\bibitem{Martinec:2023zha}
E.~J. Martinec, {\it {AdS3 Orbifolds, BTZ Black Holes, and Holography}},
  [\href{http://xxx.lanl.gov/abs/2307.02559}{{\tt arXiv:2307.02559}}].

\bibitem{Bertle:2020sgd}
H.~Bertle, A.~Dei, and M.~R. Gaberdiel, {\it {Stress-energy tensor correlators
  from the world-sheet}},  {\em JHEP} {\bf 03} (2021) 036,
  [\href{http://xxx.lanl.gov/abs/2012.08486}{{\tt arXiv:2012.08486}}].

\bibitem{Kim:2015gak}
J.~Kim and M.~Porrati, {\it {On the central charge of spacetime current
  algebras and correlators in string theory on AdS$_{3}$}},  {\em JHEP} {\bf
  05} (2015) 076, [\href{http://xxx.lanl.gov/abs/1503.07186}{{\tt
  arXiv:1503.07186}}].

\bibitem{Lee:1998bxa}
S.~Lee, S.~Minwalla, M.~Rangamani, and N.~Seiberg, {\it {Three point functions
  of chiral operators in D = 4, N=4 SYM at large N}},  {\em Adv. Theor. Math.
  Phys.} {\bf 2} (1998) 697--718,
  [\href{http://xxx.lanl.gov/abs/hep-th/9806074}{{\tt hep-th/9806074}}].

\bibitem{Freedman:1998tz}
D.~Z. Freedman, S.~D. Mathur, A.~Matusis, and L.~Rastelli, {\it {Correlation
  functions in the CFT(d) / AdS(d+1) correspondence}},  {\em Nucl. Phys. B}
  {\bf 546} (1999) 96--118, [\href{http://xxx.lanl.gov/abs/hep-th/9804058}{{\tt
  hep-th/9804058}}].

\bibitem{Harlow:2011ny}
D.~Harlow, J.~Maltz, and E.~Witten, {\it {Analytic Continuation of Liouville
  Theory}},  {\em JHEP} {\bf 12} (2011) 071,
  [\href{http://xxx.lanl.gov/abs/1108.4417}{{\tt arXiv:1108.4417}}].

\bibitem{Dei:2022pkr}
A.~Dei and L.~Eberhardt, {\it {String correlators on $\text{AdS}_3$: Analytic
  structure and dual CFT}},  {\em SciPost Phys.} {\bf 13} (2022), no.~3 053,
  [\href{http://xxx.lanl.gov/abs/2203.13264}{{\tt arXiv:2203.13264}}].

\end{thebibliography}
\end{document}